\documentclass[twocolumn,showpacs,amsmath,amssymb,pre]{revtex4}

\usepackage{graphicx}
\usepackage{dcolumn}

\begin{document}
\title{Mode-coupling theory and molecular
dynamics simulation for heat conduction in a chain with transverse motions}

\author{Jian-Sheng Wang} 
\affiliation{Singapore-MIT Alliance and Department of Computational Science,National University of Singapore, Singapore 117543, Republic of Singapore}

\author{Baowen Li}
\affiliation{
Department of Physics, National University of Singapore, Singapore 117542, Republic of Singapore}

\date{5 March 2004}

\begin{abstract}
We study heat conduction in a one-dimensional chain of particles
with longitudinal as well as transverse motions.  The particles are
connected by two-dimensional harmonic springs together with bending
angle interactions. The problem is analyzed by mode-coupling theory
and compared with molecular dynamics.  We find very good, quantitative
agreement for the damping of modes between a full mode-coupling theory
and molecular dynamics result, and a simplified mode-coupling theory
gives qualitative description of the damping.  The theories predict
generically that thermal conductance diverges as $N^{1/3}$ as the size
$N$ increases for systems terminated with heat baths at the ends.  The
$N^{2/5}$ dependence is also observed in molecular dynamics which we
attribute to crossover effect.
\end{abstract}

\pacs{44.10.+i,  05.45.--a, 05.70.Ln, 66.70.+f}
\keywords{heat conduction, 1D thermal transport, mode-coupling}
\maketitle

\section{Introduction}

The problem of heat conduction is a well-studied field.  More than two
centuries ago, Joseph Fourier summarized the behavior of heat
conduction by the law that bears his name.  This law describes
phenomenologically that the heat current is proportional to the
temperature gradient.  The detailed atomistic theories of heat
conduction appeared only much later.  For heat conduction in gas, the
simple kinetic theory gives the result $\kappa = \frac{1}{3}C\bar v
\bar l$, where $C$ is specific heat, $\bar v$ is average velocity and
$\bar l$ is mean-free path.  Peierls' theory of heat conduction in
insulating solids \cite{peierls} is a classic on this subject.  These
early theories deal with mostly the relevant three dimensions.  It
turns out that low dimensional systems are more interesting and in
some sense strange.  An analysis of a simple one-dimensional (1D) harmonic oscillator
model shows \cite{rieder-lebowitz-lieb} that there is no well-defined
temperature gradient, the thermal conductivity diverges with system
sizes as $N^{1/2}$ or $N$, depending on the boundary conditions.  There
is a general argument, for momentum conserving systems, that the
thermal conduction in 1D is necessarily divergent \cite{Prosen}.

There have been many analytical and numerical studies of 1D heat
conduction (see ref.~\cite{Lebowitz} and \cite{Lepri-review} for
review).  We'll mention some of the most relevant papers to current
work.  The work of Lepri {\sl et al} \cite{2/5} by mode-coupling
theory and molecular dynamics suggests a divergent thermal
conductivity exponent of $2/5$, i.e., $\kappa \propto N^{2/5}$ for a
1D chain model with Fermi-Pasta-Ulam (FPU) interactions.
Mode-coupling theory is usually applied in the dynamics of liquids
\cite{pomeau,balucani}.  The first use of this theory in the context
of heat conduction appears only recently, mostly due to Lepri and his
collaborators \cite{lepri-mode-coupling}.  Perverzev analyzed the same
problem with the Peierls theory for phonon gas and gave the same
conclusion of $2/5$ exponent \cite{perverzev}.  The result of $2/5$ is
also supported by numerical simulation from several groups
\cite{FPUexp-Kaburaki,FPUexp-Lepri-PRL,FPUexp-Lepri,FPUexp-Hu}.  
These results are supposed to be universal to some extent.  However,
it is challenged by a different result of $1/3$ by Narayan and
Ramaswamy \cite{1/3}, based on fluctuating hydrodynamics and
renormalization group analysis.  The numerical result for this 1/3 law is
not convincing, as for the same model -- the hard-particle gas model
-- some obtained 1/3 \cite{deutsch-narayan,GrassbergerYang}, while
others obtained different value 1/4 \cite{Casati0}.  But for the FPU
model, there is no good evidence for an exponent of 1/3
\cite{lepri-recent}.

When momentum conservation is broken such as the one with on-site
potential, the heat conduction can become normal again like the
Frenkel-Kontorova model \cite{Li98} and the $\phi^4$ model
\cite{FPUexp-Hu,phi4}.  

In order to understand the underlying microscopic dynamical mechanism
of the Fourier law, a different class of models -- billiard channels
-- have been introduced and studied in recent years
\cite{mixing,LiWang}. Various exponent values are found in such
systems. Thus, it is believed that a universal constant does not exist
at all. Instead, the divergent (convergent) exponent of the thermal
conductivity is found related to the power of super (sub) diffusion
\cite{LiWang}.

Besides the theoretical significance of heat conduction research in
low dimensional systems, it is also of practical importance.  Recent
development of nanotechnology will enable us to manufacture devices
with feature sizes at molecular level.  The understanding of heat
conduction mechanism will allow us to control and manipulate heat
current, and eventually to design novel thermal devices with certain
function \cite{diode}.  To this end, more realistic physical models
are necessary.  Among many others, nanotubes and polymer chains are
most promising.  There have been a number of numerical works to compute
the thermal conductivity of the Carbon nanotubes
\cite{carbon-nanotube-kappa-1,carbon-nanotube-kappa-2}.  
Recent molecular dynamics (MD) study of Carbon nanotubes with
realistic interaction potential suggested a divergent thermal
conductivity for narrow diameter tubes \cite{Nanotube,NUSpaper}.  The quantum
effect of such systems is also very interesting
\cite{quantum-kappa,quantum-kappa-Japan}.

We study the heat conduction of a 1D solid, as a classical system.  A
brief version of this paper is reported in ref.~\cite{Wang-Li}.  In
the rest of the paper, we introduce the quasi-one-dimensional chain
model in Sec.~\ref{sec:chain}.  We discuss the basis of the
mode-coupling theory, the projection method in
Sec.~\ref{sec:projection}.  The mode-coupling approximations and their
numerical and analytical solutions are discussed in
Sec.~\ref{sec:mode-coupling} and \ref{sec:solution}.  The basic output
of the mode-coupling analysis is the dependence of damping of the
modes with the wave-vector of the modes.  We find that the transverse
modes are diffusive, with $\gamma_p^\perp \propto p^2$, while the
longitudinal modes are super-diffusive, $\gamma_p^\parallel \propto
p^{3/2}$, where $\gamma_p$ is the decay rate for mode with momentum or
lattice wave number $p$.  We discuss the relationship between the
damping of the modes with the heat conductivity through Green-Kubo
formula in Sec.~\ref{sec:green-kubo}.  Our mode-coupling theory
predicts that the heat conductance diverges with the $1/3$ exponent
when the transverse motions are important, while $2/5$ is recovered if
the transverse motions can be neglected.  In Sec.~\ref{sec:MD}, we
present nonequilibrium molecular dynamics results (with heat baths) of
heat conductance and compare with mode-coupling theory.  We conclude
in the last section.

\section{\label{sec:chain}Chain Model}

Most of the previous studies considered only strictly 1D models, with
the Fermi-Pasta-Ulam (FPU) model as the most representative. The strictly
1D models may not be applicable to real systems such as the nanotubes.
Real systems of nanotubes or wires live in three-dimensional space.
The added transverse motion and the flexibility of the tube at long
length scales will certainly scatter phonons, and thus should have a
profound effect on thermal transport.

While a direct simulation of a realistic system such as a polyethelene
chain with empirical force fields such as that in
refs.~\cite{goddard,Manevitch} or nanotube with Tersoff potential
\cite{tersof} is possible, we think it is useful to consider a
simplified model which captures one of the important features of the
real systems -- transverse degrees of freedom.  Therefore we propose to
study the following chain model in two dimensions \cite{Wang-Li}:
\begin{eqnarray}
H({\bf p}, {\bf r}) & = & \sum_{i} \frac{{\bf p}_i^2}{2m} +   
\frac{1}{2} K_r \sum_{i} \Bigl( | {\bf r}_{i+1}-{\bf r}_i | -a\Bigr)^2 
\nonumber
\\
&& +\, K_\phi \sum_{i} \cos \phi_i,
\end{eqnarray} 
where the position vector ${\bf r}=(x,y)$ and momentum vector ${\bf p}
=(p_x, p_y)$ are two-dimensional; $a$ is lattice constant.  The
minimum energy state is at $(ia,0)$ for $i=0$ to $N-1$.  If the system
is restricted to $y_i=0$ (corresponding to $K_\phi=+\infty$), it is
essentially a 1D gas with harmonic interaction.  The coupling $K_r$ is
the spring constant; $K_\phi$ signifies bending or flexibility of the
chain, while $\phi_i$ is the bond angle formed with two neighboring
sites, $\cos \phi_i = - {\bf n}_{i-1} \cdot {\bf n}_{i}$, and unit
vector ${\bf n}_i = \Delta {\bf r}_i /| \Delta {\bf r}_i|$, $\Delta
{\bf r}_i = {\bf r}_{i+1} - {\bf r}_{i}$.

Unlike the FPU model which does not have an energy scale, the second
bond-angle bending term introduces an energy scale.  In this work, we
take mass $m=1$, spring constant $K_r=1$, and the Boltzmann constant
$k_B=1$, thus the most important parameters are $K_\phi$ and
temperature $T$.

\section{\label{sec:projection}Projection Method}
\subsection{Basic theory of projection}
We follow the formulation of the projection method in
ref.~\cite{Kubo}.  Let
\begin{equation}
A = \begin{pmatrix} 
           a_1 \cr
           a_2 \cr
           \vdots \cr
           a_n \cr 
     \end{pmatrix}
\end{equation}
be a column vector of $n$ components of some arbitrary functions of
dynamical variables $(p,q)$.  Each of the function $a_j(p,q)$ can be
complex.  Later, we shall choose $a_j$ to be the canonical coordinates
of the system. We use $A^\dagger = (a_1^*, a_2^*, \cdots, a_n^*)$ to
denote the Hermitian conjugate of $A$.  The equation of motion for $A$
is
\begin{equation}
 \dot A_t = {\cal L} A_t,  \quad \mbox{or}\quad
\dot a_j(t) = {\cal L} a_j(t),
\label{Eqmotion}
\end{equation}
where $\cal L$ is the Liouville operator
\begin{equation}
{\cal L} = - \sum { \partial H \over \partial q } { \partial \over \partial p}
 + \sum { \partial H \over \partial p } { \partial \over \partial q}.
\end{equation}
Equation~(\ref{Eqmotion}) can be viewed as a partial differential
equation with variables $(p,q)$ and time $t$.  $A_t \equiv A(p_t, q_t) =
A(t,p,q)$, i.e., the quantity $A$ at time $t$ when the initial
condition at $t=0$ is $(p,q)$.  Quantity without a subscript $t$ will
be understood to be evaluated at time $t=0$, e.g., $p=p_{t=0} = p(0)$.
A formal solution to Eq.~(\ref{Eqmotion}) is simply $A_t = e^{t {\cal
L}} A(p,q)$.

The projection operator on a column vector $X$ is defined by
\begin{equation}
{\cal P} X = \langle X, A^\dagger \rangle 
             \langle A, A^\dagger \rangle^{-1} A,
\label{projector}
\end{equation}
where $\langle X, A^\dagger \rangle$ and $\langle A, A^\dagger
\rangle$ are $n \times n$ matrices.  The angular brackets denote
thermodynamical average in a canonical ensemble at temperature $T$.
The comma separating the two terms is immaterial, but we use a
notation of inner product.  One can verify that $\cal P$ is indeed a
projection operator, i.e., ${\cal P}^2 = {\cal P}$.

If we apply the projection operator ${\cal P}$ and ${\cal P}' =
1-{\cal P}$ to the equation of motion, we get two coupled equations.
Solving formally the second equation associated with ${\cal P}'$, and
substituting it back into the first equation, we obtain an equation
for the projected variable that formally resembles a Langevin
equation:
\begin{equation}
\dot A_t = i \Omega A_t - \int_0^t \Gamma(t-s) A_s\, ds + R_t,
\label{Langevin}
\end{equation}
where $i=\sqrt{-1}$ is the complex unit, and
\begin{eqnarray}
 \Gamma(t) &=& \langle R_t, R_0^\dagger \rangle 
             \langle A, A^\dagger \rangle^{-1},\\
 i\Omega &=& \langle  \dot A, A^\dagger \rangle
             \langle  A, A^\dagger \rangle^{-1},\\
 R_t &=& e^{t {\cal P}' {\cal L}} R_0,  \quad
 R_0 = \dot A - i \Omega A = {\cal P}' {\cal L} A.
\end{eqnarray}
This set of equations is deterministic and formally exact.  The only
assumption made is that equilibrium distribution can be realized.

What is more important is the correlation functions, which are
physical observables.  We define the normalized correlation function
(correlation matrix) as
\begin{equation}
G(t) = \langle A_t, A_0^\dagger \rangle 
       \langle A, A^\dagger \rangle^{-1}. 
\end{equation}
It is an identity matrix at $t=0$ and has the property that $ \dot
G(0) = i \Omega$.  From Eq.~(\ref{Langevin}) we have,
\begin{equation}
\dot G(t) = i \Omega G(t) - \int_0^t \Gamma(t-s) G(s)\, ds.
\end{equation}
This equation can be solved formally using proper initial condition
with a Fourier-Laplace transform,
\begin{equation}
G[z] = \int_0^\infty e^{-i z t} G(t)\, dt,
\end{equation}
and similarly for $\Gamma(t)$. The solution is 
\begin{equation}
G[z] = \bigl( i(z - \Omega) + \Gamma[z] \bigr)^{-1}.
\label{EqGw}
\end{equation}
To simplify notation, we have used parentheses (e.g., $G(t)$) for
functions in time domain, and square brackets with the same symbol
(e.g., $G[z]$) for their corresponding Fourier-Laplace transform in
frequency domain.

The information about the system is in the memory kernel $\Gamma(t)$.
However, such a correlation function is difficult to calculate, since
the evolution of the ``random force'' $R_t$ does not follow the
dynamics of the original Hamiltonian system.  For example, it is
impossible to compute directly $R_t$ from molecular dynamics.  For
this reason, a ``true force'' can be introduced which obeys the normal
evolution, i.e.,
\begin{equation}
F_t = e^{t {\cal L}} R_0 =  \dot A_t - i \Omega A_t.
\end{equation}
The correlation function of the true force,
\begin{equation}
\Gamma_F(t) = \langle F_t, F_0^\dagger  \rangle
              \langle A, A^\dagger \rangle^{-1},
\end{equation}
and that of the random force are related in Fourier space as \cite{Kubo}
\begin{equation}
\Gamma[z]^{-1} = \Gamma_F[z]^{-1} - 
                      \bigl( i (z - \Omega) \bigr)^{-1}.
\label{EqF2R}
\end{equation}
This completes the formal theory of projection due originally to
Zwanzig \cite{Zwanzig} and Mori \cite{Mori}.  These results are formal
and exact.  They give us relation between correlation of the ``force''
and correlation of dynamical variables.  They are the starting point
for mode-coupling theory.  In the next subsections, we apply it to our
chain model and introduce a series of approximations to solve it.

\subsection{The chain model}
We now apply the projection method to our chain model.  We choose the
normal modes as the basic quantities $a_j$ that we are going to
project out.  There are several reasons for choosing the normal modes.
To zeroth order approximation, each mode is nearly independent.  The
slowest process corresponds to long wave-length modes.  The effect of
short wave-length modes can be treated as stochastic noise (the random
force $R_t$).

We choose $A$ to be the complete set of canonical momenta and
coordinates:
\begin{equation}
A = \begin{pmatrix} 
           P_k^\parallel \cr
           P_k^\perp \cr
           Q_k^\parallel \cr
           Q_k^\perp \cr 
     \end{pmatrix}, \quad k = 0, 1, \cdots, N-1,
\end{equation}
where
\begin{eqnarray}
Q_k^\parallel &=& \sqrt{\frac{m}{N}} \sum_{j=0}^{N-1} u_j e^{i2\pi j k/N},\\
Q_k^\perp &=& \sqrt{\frac{m}{N}} \sum_{j=0}^{N-1} y_j e^{i2\pi j k/N},\\
P_k^\parallel &=& \dot Q_k^\parallel = \frac{1}{\sqrt{mN}} 
                  \sum_{j=0}^{N-1} p_{j,x} e^{i2\pi j k/N},\\
P_k^\perp &=& \dot Q_k^\perp = \frac{1}{\sqrt{mN}} 
                  \sum_{j=0}^{N-1} p_{j,y} e^{i2\pi j k/N}.
\end{eqnarray}
We have defined the position vector ${\bf r}_j = (x_j, y_j) = (u_j +
aj, y_j)$, so that $u_j$ and $y_j$ are deviations from
zero-temperature equilibrium position.  Because the Fourier transform
is a periodic function, the index $k$ is unique only modulo $N$.  As a
result, we can also let $k$ vary in the range $- N/2 \leq k < N/2$.
We also note that $Q_k^* = Q_{-k}$.

With these definitions, we can compute the matrix $\Omega$ and
expression for the true force $F$ in the general theory.  We find that
for $\langle A, A^\dagger \rangle$, the components are
\begin{eqnarray}  
\langle P_k^{\mu} P_{k'}^{\nu}{}^* \rangle &=& 
                     \delta_{kk'} \delta_{\mu \nu}\frac{1}{\beta}, 
       \quad \mu,\nu = \,\parallel, \perp,\\
\langle P_k^{\mu} Q_{k'}^{\nu}{}^* \rangle &=& 0,\\
\langle Q_k^{\mu} Q_{k'}^{\nu}{}^* \rangle &=& 
                   \delta_{kk'} \delta_{\mu \nu}
                  \frac{1}{\beta (\tilde \omega_k^{\mu})^2},\quad 
                  \beta = \frac{1}{k_B T}.
\end{eqnarray}
We have used equal-partition theorem for the average kinetic energy
expression, and the last equation merely defines the effective
frequencies $\tilde \omega_k^{\mu}$ for each mode.  Note that due to
translational invariance, correlation between different $k$ modes
vanishes.  Correlation between transverse and longitudinal modes also
vanishes due to the reflection symmetry of $y_j \to -y_j$ for the chain.
Thus equal-time correlation for $A$ is diagonal,
\begin{equation}
\langle A, A^\dagger \rangle = \frac{1}{\beta} \begin{pmatrix}
                                I  & 0 \cr
                                0  & \tilde \omega^{-2} \cr
                               \end{pmatrix}.
\end{equation}
We have defined $\tilde \omega$ as a $2N \times 2N$ diagonal matrix
with elements $\tilde \omega_k^\mu$; $I$ is a $2N\times 2N$ identity
matrix.  Similarly, the correlation $\langle \dot A, A^\dagger
\rangle$ is found from
\begin{eqnarray}  
\langle \dot P_k^{\mu} P_{k'}^{\nu}{}^* \rangle &=& 0,\\ 
\langle \dot P_k^{\mu} Q_{k'}^{\nu}{}^* \rangle &=&  - 
                     \delta_{kk'} \delta_{\mu \nu}\frac{1}{\beta}, \\
\langle \dot Q_k^{\mu} P_{k'}^{\nu}{}^* \rangle &=& 
                     \delta_{kk'} \delta_{\mu \nu}\frac{1}{\beta}, \\
\langle \dot Q_k^{\mu} Q_{k'}^{\nu}{}^* \rangle &=& 0.
\end{eqnarray}
The second equation is from a general virial theorem \cite{Huang}.
We have
\begin{equation}  
i \Omega = \langle \dot A, A^\dagger \rangle
\langle A, A^\dagger \rangle^{-1} = \begin{pmatrix}
                                     0 & -\tilde \omega^2 \cr
                                     I  &    0 \cr
                                    \end{pmatrix}.
\end{equation}
The expression for the true force $F = \dot A - i \Omega A$ is then
\begin{equation}
F \equiv 
\begin{pmatrix} 
        F^P \cr 
        F^Q \cr
\end{pmatrix} = \begin{pmatrix}
       \dot P_k^\mu + \tilde \omega_k^\mu{}^2 Q_k^\mu \cr
        0 \cr
\end{pmatrix}.
\end{equation}
Note that only the momentum sector has a nonzero value, and $\dot
P_k^\mu = \ddot Q_k^\mu$ is given approximately by
Eq.~(\ref{Eqmotion1}) and (\ref{Eqmotion2}) below.  With this special
form of $F$, the damping matrix $\Gamma$ is also diagonal and is
nonzero only in the $PP$ components.  With these results,
Eq.~(\ref{EqGw}) becomes
\begin{equation}
 G[z]  = \begin{pmatrix}
               iz d  &  - \tilde \omega^2 d \cr
               d          &  (iz + \tilde \Gamma[z]) d \cr
              \end{pmatrix},
\label{general-Gz}
\end{equation}
where $d = ( \tilde \omega^2 - z^2 I + iz \tilde \Gamma[z])^{-1}$ is a
$2N \times 2N$ diagonal matrix, and $\tilde \Gamma[z]$ is the
Fourier-Laplace transform of the correlation $\beta\langle R_t^P,
(R^P)^\dagger \rangle$, which is also diagonal (we'll denote as
$\Gamma_{k}^\mu(t)$).  In particular, we have the usual expression for
the normalized coordinate correlation,
\begin{eqnarray}
g_{QQ,k}^\mu(t) &=& { \langle Q_k^\mu(t) Q_{-k}^\mu(0) \rangle \over
                  \langle |Q_k^\mu|^2 \rangle },\\
g_{QQ,k}^\mu[z] &=& { i z + \Gamma_k^\mu[z] \over
            \tilde \omega_k^\mu{}^2 - z^2 + iz \Gamma_k^\mu[z]}.
\label{Eqvtog}
\end{eqnarray}
For simplicity, we drop the $QQ$ subscript for the coordinate
correlation for the rest of the paper.

Finally, the relation between the random force correlation and true
force correlation, Eq.~(\ref{EqF2R}), becomes,
\begin{equation}
 \frac{1}{\Gamma_k^\mu[z]} = \frac{1}{\Gamma_{F,k}^\mu[z]}  
    - {i z \over \tilde \omega_k^\mu{}^2 - z^2 }.
 \label{EqvFtov}
\end{equation} 
The force correlation $\Gamma_{F,k}^\mu$ can be computed from the
correlation function $\beta\langle F_t^P, (F^P)^\dagger \rangle$.  It
is convenient to separate the linear term in the force from the
nonlinear contribution.  So we write
\begin{equation}
F^P(t) = \tilde\omega^2 Q(t) + \ddot Q(t)  =
(\tilde\omega^2 - \omega_0^2) Q(t) + f_N(t), 
\end{equation}
where $\tilde\omega$ is effective angular frequency and $\omega_0$ is
bare angular frequency of the mode.  We have dropped the indices $k$
and $\mu$ since these equations apply for any of the modes.  $f_N$ is
at least quadratic in $Q$.  We note that the correlation of $F^P(t)$
is linearly related to the coordinate correlation function $g[z]$,
\begin{equation}
\Gamma_F[z] = { - iz(\tilde\omega^2-z^2) \over \tilde\omega^2 } 
+ { (\tilde\omega^2 - z^2)^2 \over \tilde\omega^2 } g[z].
\label{EqgtoGamF}
\end{equation}
This is simply a consequence of Eqs.~(\ref{Eqvtog}) and
(\ref{EqvFtov}), but can also be derived directly from the definition.
The second derivative of $Q(t)$ in frequency domain is $Q[z]$ multiplied by
$(iz)^2$.  Using the fact that
\begin{equation}
\lim_{\epsilon \to 0^+}\int_0^\infty \ddot Q(t) e^{-izt - \epsilon t} dt 
= -z^2 Q[z] - \dot Q(0) - iz Q(0),
\end{equation}
we can also derive Eq.~(\ref{EqgtoGamF}) with the understanding that
$\langle \cdots \rangle$ is an average over the initial conditions.
Since $g[z]$ is finite or at least should not diverge precisely at
$z=\tilde \omega$, this implies $\Gamma_F[\tilde \omega] = 0$.

With the above results, we can derive an expression of the true force
correlation in terms of nonlinear force correlation,
\begin{eqnarray}
{ \Gamma_F(t)\over \beta} &=& \langle F^P(t) F^P(0)^* \rangle \nonumber \\
 &=& \Delta \tilde \omega^4 \langle Q(t)Q^*(0) \rangle 
     +  \langle f_N(t) f_N(0)^* \rangle  \nonumber \\
 &&    - \Delta \tilde\omega^2 \Bigl( 
       \langle f_N(t) Q^*(0) \rangle + \langle Q(t) f_N^*(0) \rangle \Bigr), 
\end{eqnarray}
where $\Delta \tilde{\omega}^2 = \omega_0^2 - \tilde{\omega}^2$.  The
mixed term can be expressed in terms of $g[z]$ by noting that $f_N =
\ddot Q + \omega_0^2 Q$.  The two mixed terms $\langle f_N(t)
Q^*(0)\rangle$ and $\langle Q(t)f_N^*(0)\rangle$ are equal due to
time-reversal symmetry.  We find
\begin{equation}
\beta \tilde\omega^2 \int_0^\infty \langle f_N(t) Q^*(0)\rangle e^{-izt} dt = 
(\omega_0^2-z^2) g[z] - iz.
\end{equation}
Finally, we have
\begin{equation}
\Gamma_F[z] = 
{\Delta \tilde\omega^2 \over \tilde\omega^2} 
\Bigl((2z^2-\omega_0^2-\tilde\omega^2)g[z]  + 2iz \Bigr)
+ \Gamma_N[z].
\end{equation}
We can also express $g[z]$ in terms of the nonlinear part of the force
correlation, $\Gamma_N$ ($=\beta \langle f_N(t) f_N(0)^*\rangle$):
\begin{equation}
 g[z] = { iz(2\omega_0^2 - \tilde\omega^2-z^2) + \tilde\omega^2 \Gamma_N[z]
   \over (z^2-\omega_0^2)^2 }.
\label{EqvNtog}
\end{equation}
Again, since $g[z]$ cannot diverge precisely at $z=\omega_0$, this
implies that $\Gamma_N[z]$ must take a special form to cancel the
apparent divergence.  Thus, if we do not take care of these
superficial divergences, we will not be able to make correct
prediction for the correlation function.

We can also relate the original damping function $\Gamma$ to the
nonlinear one,
\begin{equation}
\Gamma[z] = { -i(\tilde \omega^2 - \omega_0^2)^2 z + 
              \tilde\omega^2 ( \tilde\omega^2 - z^2) \Gamma_N[z]
      \over \omega_0^4 - \tilde \omega^2 z^2 - iz\tilde \omega^2 \Gamma_N[z]}.
\label{EqvNtov}
\end{equation}
This last equation is useful for approximating the damping function.
All of these relations are exact.  This ends our formal application of
the projection method to the chain model.
         
\section{\label{sec:mode-coupling}Mode-Coupling Theory}

To make some progress for analytic and numerical treatment, we have to
make some approximations.  First, we'll consider small oscillations
valid at relatively low temperatures.  An approximate Hamiltonian for
small oscillations near zero-temperature equilibrium position, keeping
only leading cubic non-linearity in the Hamiltonian, is then given by
\begin{eqnarray}
H(P,Q) &=& {1 \over 2} \sum_{k; \mu=\parallel,\perp} \bigl( P_k^\mu P_{-k}^\mu 
  + \omega_k^\mu{}^2 Q_k^\mu Q_{-k}^\mu \bigr) \nonumber \\
  && + \!\!\!\!\!\sum_{k+p+q \equiv 0 \bmod N} 
       \!\!\!\!\!c_{k,p,q} Q_k^\parallel Q_p^\perp Q_q^\perp,
\label{H-perturb}
\end{eqnarray}
where  
\begin{eqnarray}
\omega_k^\parallel{}^2 &=& { 4 K_r \over m} \sin^2 {k\pi \over N},
\label{bare-1}\\
\omega_k^\perp{}^2 &=& { 16 K_\phi \over m a^2} \sin^4 {k\pi \over N},
\label{bare-2}
\end{eqnarray}
are the `bare' dispersion relations, and 
\begin{eqnarray}
c_{k,p,q} &=& { 8 i \over a^3 m^{3/2} N^{1/2} } 
           \sin{k\pi \over N} \sin{p\pi \over N} \sin{q\pi \over N} 
           \Bigl( \frac{1}{2} a^2 K_r \nonumber \\
        && + K_\phi (-2  
           + \cos{2\pi p\over N} + \cos { 2 \pi q \over N}) \Bigr).
\end{eqnarray}
The absence of $Q^\parallel Q^\parallel Q^\parallel$ term in Hamiltonian (\ref{H-perturb}) is due to the
quadratic nature of the potential, while the absence of the terms of the
form $Q^\perp Q^\perp Q^\perp$ and $Q^\parallel Q^\parallel Q^\perp$
is due to the reflective symmetry about $y$-axis of the Hamiltonian.
We view $k$ and $-k$ as independent component when taking the
derivatives.  A slightly modified Hamilton's equation (because of the
use of complex numbers) describes the dynamics:
\begin{eqnarray}
\dot P_k^\nu &=& - { \partial H \over \partial Q_{-k}^\nu},\\
\dot Q_k^\nu &=& {\partial H \over \partial P_{-k}^\nu}.
\end{eqnarray}
This gives the following equations of motion:
\begin{eqnarray}
\ddot Q_k^\parallel &=& - \omega_k^\parallel{}^2 Q_k^\parallel + 
          \sum_{k'+k''=k} c_{k',k''}^{YY} Q_{k'}^\perp Q_{k''}^\perp,
          \label{Eqmotion1}\\
\ddot Q_k^\perp &=& - \omega_k^\perp{}^2 Q_k^\perp + 
          \sum_{k'+k''=k} c_{k',k''}^{UY} Q_{k'}^\parallel Q_{k''}^\perp,
          \label{Eqmotion2}
\end{eqnarray}
where 
\begin{eqnarray}
c_{k,p}^{YY} &&= i 4 \sqrt{\frac{1}{N m^3 a^2}} 
        \sin\frac{k\pi}{N} \sin\frac{p\pi}{N} \sin\frac{(k+p)\pi}{N}\nonumber \\
             && \Bigl( K_r + \frac{2}{a^2} K_\phi \bigl( -2 +           
             \cos \frac{2k\pi}{N} + \cos\frac{2p\pi}{N} \bigl) \Bigl),\\
c_{k,p}^{UY} &&= i 8 \sqrt{\frac{1}{N m^3 a^2}} 
        \sin\frac{k\pi}{N} \sin\frac{p\pi}{N} \sin\frac{(k+p)\pi}{N} 
            \Bigl( K_r \nonumber \\
             && + \frac{2}{a^2} K_\phi \bigl( -2 +           
             \cos \frac{2p\pi}{N} + \cos\frac{2(k+p)\pi}{N} \bigl) \Bigl).
\end{eqnarray}
With these expressions, we are ready to compute the force correlation
function in terms of dynamic variables $Q_k^\mu$.  We write
\begin{equation}
F_{k,\mu}^P = - \Delta \omega_k^\mu{}^2  Q_k^\mu + f_{k,\mu}^N,
\end{equation}
where $\Delta \omega_k^\mu{}^2 = \omega_k^{\mu}{}^2 - \tilde
\omega_k^\mu{}^2$ is the difference between bare dispersion
relation and effective dispersion relation. The second term
$f_{k,\mu}^N$ denotes the rest of the nonlinear force (we take only
the quadratic terms in $Q$).  Due to translational and reflective
symmetries, the correlation matrix formed by $F^P$ is diagonal without
any approximation. The time-displaced correlation for the diagonal
terms defines true force correlation.  The nonlinear part of the
contributions is
\begin{eqnarray} 
\Gamma_{N,k}^\parallel(t) \approx  
 \beta \sum_{k_1+k_2=k} \sum_{k_3+k_4=k} \qquad\qquad\qquad\qquad\qquad \nonumber \\
  c_{k_1,k_2}^{YY} c_{k_3,k_4}^{YY*} 
\langle Q_{k_1}^\perp(t) Q_{k_2}^\perp(t) Q_{k_3}^{\perp*}(0) 
Q_{k_4}^{\perp*}(0) \rangle, \quad{} \\
\Gamma_{N,k}^\perp(t) \approx  
 \beta \sum_{k_1+k_2=k} \sum_{k_3+k_4=k} \qquad\qquad\qquad\qquad\qquad \nonumber \\
\!\!\!\!\!c_{k_1,k_2}^{UY} c_{k_3,k_4}^{UY*} 
\langle Q_{k_1}^\parallel(t) Q_{k_2}^\perp(t) Q_{k_3}^{\parallel*}(0) 
Q_{k_4}^{\perp*}(0) \rangle.\quad{}
\end{eqnarray}

In order to have a closed system of equations for the normalized
correlation functions, we use the standard mean-field type
approximation, $\langle Q Q Q Q \rangle \approx \langle Q Q \rangle
\langle Q Q \rangle$.  Owing to the $\delta$ correlation between
different $k$, the double summation can be reduced to a single one.
We introduce
\begin{equation}
\nu_{N,k}^\mu(t) = \Gamma_{N,k}^\mu(t)/p^2, \qquad p = \frac{2\pi k}{N a},
\label{EqGammatonu}
\end{equation} 
and similarly $\nu_{k}^\mu(t)$ associated with $\Gamma_k^\mu$.  In
terms of $\nu_N(t)$, we obtain
\begin{eqnarray}
\nu_{N,k}^\parallel(t) &=& 
\!\!\!\sum_{k_1+k_2=k}\!\!\!\! K_{k_1,k_2}^\parallel g_{k_1}^\perp(t) g_{k_2}^\perp(t),\label{Eqmodekp-full-para}\\
\nu_{N,k}^\perp(t) &=& 
\!\!\!\sum_{k_1+k_2=k}\!\!\!\! K_{k_1,k_2}^\perp g_{k_1}^\parallel(t) g_{k_2}^\perp(t), \label{Eqmodekp-full-perp}
\end{eqnarray}
where
\begin{eqnarray}
K_{k_{1},k_{2}}^\parallel &=& 2k_B T \left| {c_{k_1,k_2}^{YY} \over 
\frac{2\pi (k_1+k_2)}{Na}
 \tilde \omega_{k_1}^\perp \tilde \omega_{k_2}^\perp }  \right|^2, 
\label{EqKpara} \\
K_{k_1,k_2}^\perp &=& k_B T \left| {c_{k_1,k_2}^{UY} \over 
\frac{2\pi (k_1+k_2)}{Na}
 \tilde \omega_{k_1}^\parallel\tilde \omega_{k_2}^\perp }  \right|^2.
\label{EqKperp}
\end{eqnarray}
Equations (\ref{Eqmodekp-full-para}) and (\ref{Eqmodekp-full-perp})
together with the relations among $\nu_{N,k}^\mu$, $\Gamma_k^\mu$, and
$g_k^\mu$, Eqs.~(\ref{Eqvtog}), (\ref{EqvNtov}), and
(\ref{EqGammatonu}), form a system of close equations, which can be
solved in principle.  However, because of the singular nature at
$z=\omega_0$ in Eq.~(\ref{EqvNtog}), any approximation to $\Gamma_N$
will destroy a subtle cancellation of the singularity, rendering the
problem impossible to solve.

\begin{figure}
\includegraphics[width=\columnwidth]{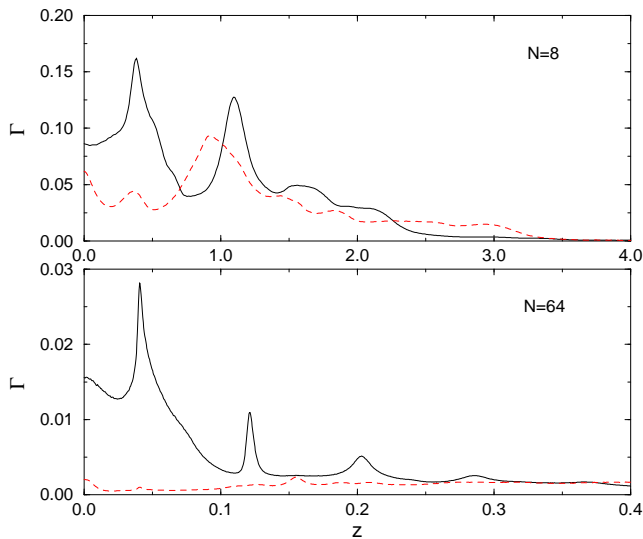}
\caption{\label{fig-MDgm} Damping functions of the slowest modes, i.e.,
real part of $\Gamma_1^\parallel[z]$ (solid line) and
$\Gamma_1^\perp[z]$ (dashed line) vs $z$, computed from molecular
dynamics for parameters of set E: $K_r=1$, $K_\phi=0.3$, $a=2$,
$T=0.4$, $m=1$.  The top figure is for size $N=8$ and bottom figure
for $N=64$.}
\end{figure}

The damping function $\Gamma_{k}^\mu[z]$ is the central function that
a successful theory needs to be able to calculate.  In
Fig.~\ref{fig-MDgm} we present examples of such functions determined
from equilibrium molecular dynamics (MD) simulation in a
microcanonical ensemble with periodic boundary condition.  A more
correct comparison of MD with mode-coupling theory should use an
ensemble of initial conditions distributed according to canonical
weight.  This may be unimportant when $N$ is large.  We compute
$\Gamma[z]$ from the ratio of two correlation functions,
$g_{QQ}[z]/g_{QP}[z] = iz + \Gamma[z]$.  For small systems, there are
a lot of peak structures associated with the low frequency modes; the
feature appears washed out for large systems.

\section{\label{sec:solution}Solution of Mode-Couping Theory}

\subsection{Numerical solution at finite $N$}

To make the problem tractable, we make a bold approximation.  Instead
of Eq.~(\ref{EqvNtov}), we take
\begin{equation}
\nu_k^\mu[z] \approx \nu_{N,k}^\mu[z].
\label{our-approximation}
\end{equation}
This is equivalent to say $\tilde \omega \approx \omega_0$.  This
appears justified for the longitudinal modes at sufficiently low
temperatures but problematic for the transverse mode, as $\tilde
\omega$ is linear in wave number $q$ but $\omega_0$ is quadratic in $q$.  
We can consider the limit of small $q$.  In such limit, the difference
between $\Gamma_F$ and $\Gamma$ is dropped.  Lepri's treatment
\cite{lepri-mode-coupling} is similar to the above approximation.  We
also note that in the work of Scheipers and Schirmacher for damping of
anharmonic crystals \cite{scheipers}, effective cubic coupling is used
instead of the `bare' coupling, $c_{k,p,q}$.  The standard operation
of projecting the random force onto bilinear form $Q_p Q_q$ to get
the mode-coupling equations also agrees with the approximation,
Eq.~(\ref{our-approximation}), but with somewhat different ``vertex
functions'' replacing Eq.~(\ref{EqKpara}) and (\ref{EqKperp}).

In order to obtain a numerical solution, besides the model parameters
(mass $m$, lattice spacing $a$, couplings $K_r$ and $K_\phi$), we also
need the effective dispersion relation.  We used MD data for this
purpose.  It turns out that a two-parameter fit of the form
\begin{equation}
\tilde \omega_k = \frac{2c}{a} \bigl| \sin\frac{k\pi}{N} \bigr|
 +\left(\omega_{{\rm max}} - \frac{2c}{a}\right)
\sin^2\frac{k\pi}{N}
\label{omega_tilde_fit}
\end{equation}
characterizes the effective dispersion relation very well, where $c$
is sound velocity at small $k$ and $\omega_{{\rm max}}$ is maximum
frequency at $k=N/2$.

We used a fast Fourier transform to solve the equations iteratively.
Given an initial $\nu[z]$, correlation function in frequency domain is
calculated by Eq.~(\ref{Eqvtog}).  Inverse transform gives us $g(t)$,
which is used to calculate the discrete sum in
Eqs.~(\ref{Eqmodekp-full-para}) and (\ref{Eqmodekp-full-perp}) to
obtain $\nu(t)$. We transform back to frequency domain for the next
iteration.

\begin{figure}
\includegraphics[width=\columnwidth]{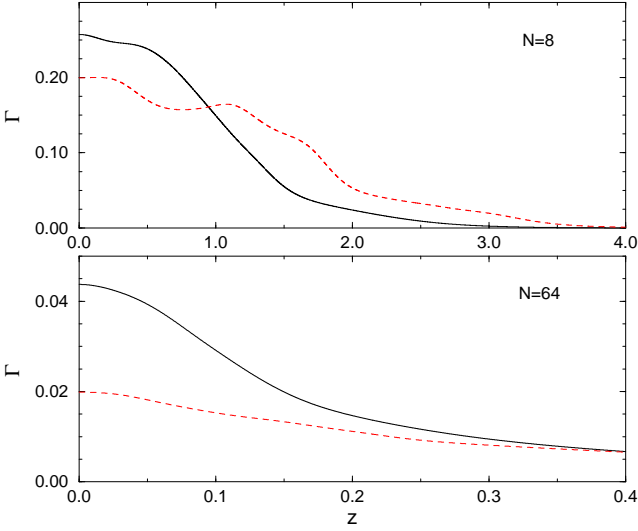}
\caption{\label{fig-MKgm} Real part of $\Gamma_1^\parallel[z]$ (solid
line) and $\Gamma_1^\perp[z]$ (dashed line) vs $z$ from a full
mode-coupling theory for parameters of set E (same as that in
Fig.\protect\ref{fig-MDgm}) at $N=8$ and 64.  The input effective
dispersion relation parameters, $c^\parallel, c^\perp, \omega_{{\rm
max}}^\parallel$, and $\omega_{{\rm max}}^\perp$, are given in
Table~\ref{tb-1}.  }
\end{figure}

\begin{figure}
\includegraphics[width=\columnwidth]{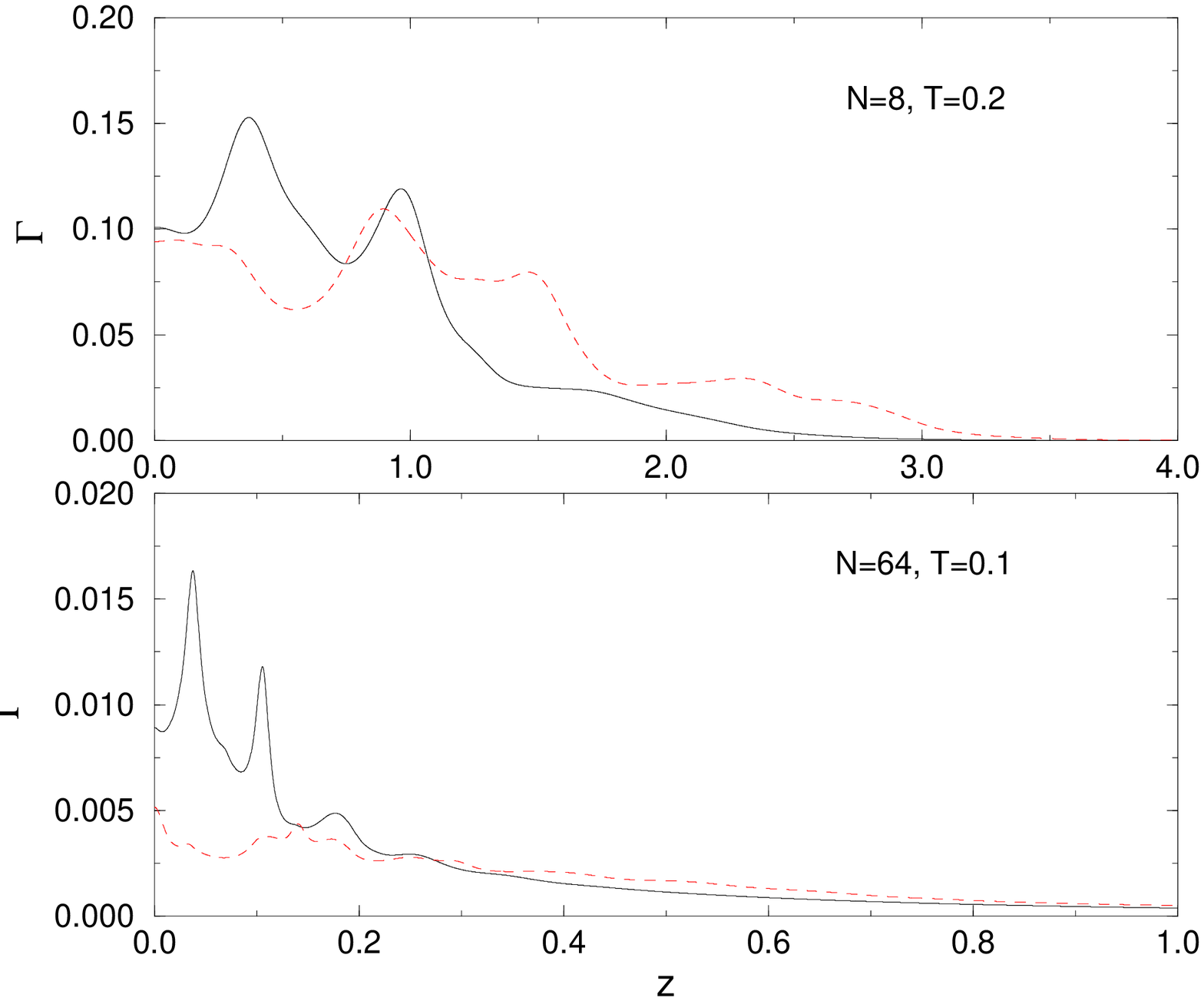}
\caption{\label{fig-MKgmh} Real part of $\Gamma_1^\parallel[z]$ (solid
line) and $\Gamma_1^\perp[z]$ (dashed line) vs $z$ from a full
mode-coupling theory for parameters of set E, but at different
temperature $T$.  Other parameters are the same as in
Fig.~\protect\ref{fig-MKgm}.  }
\end{figure}

Figure~\ref{fig-MKgm} is our theoretical calculation with parameters
exactly at set E.  This figure should be compared with the MD data in
Fig.~\ref{fig-MDgm}.  We observe that the predicted results are too
smooth without much structure.  The mode-coupling theory predicts a
damping that is a factor of two larger for the longitudinal modes and
much too large for the transverse modes.  By adjusting the temperature
to a lower value (one half for $N=8$ and a quarter for $N=64$), we
obtain curves which closely resemble the MD results.  However, the
mode-coupling theory still overestimates the value of transverse
damping.  It is clear that $K^\perp$ is considerably smaller in
actually system, but our mode-coupling theory can not produce a small
enough $K^\perp$.
 
\subsection{An effective Hamiltonian approach}

The results presented in Fig.~\ref{fig-MKgmh} give us some hints that
the mode-coupling equations are essentially correct, but the
parameters of the model need to be adjusted.  In fact, simple
perturbative expansion of the Hamiltonian can not correctly predict
$\tilde \omega^\perp_k$.  According to MD result, $\tilde
\omega^\perp_k \propto k$, but the leading contribution from
perturbation calculation should give the `bare' frequency, $\tilde
\omega^\perp_k = \omega^\perp_k \propto k^2$.

The adjusting of the parameters can be made more rigorous
\cite{scheipers} with an introduction of an effective Hamiltonian,
\begin{eqnarray}
&& \tilde H(P,Q) = \frac{1}{2} \sum_{k,\mu=\parallel,\perp} \Bigl(
P^\mu_k P^\mu_{-k} + (\tilde \omega^\mu_k)^2 Q^\mu_k Q^\mu_{-k} \Bigr) 
\nonumber \\
&& \quad + \sum_{k+p+q=0} {\tilde V}_{kpq} Q^\parallel_k Q^\perp_p Q^\perp_q
    + \sum_{k+p+q=0} {\tilde V}^{(3)}_{kpq} Q^\parallel_k 
                      Q^\parallel_p Q^\parallel_q  \nonumber \\
&& \quad =  H^0 + H'. 
\end{eqnarray}
The form of the Hamiltonian is dictated by symmetry.  Translational
invariance requires that $k+p+q=0 \pmod N$.  The system is symmetric
under reflection about $y$-axis or $x$-axis.  Thus the Hamiltonian
should be invariant under the transformation $Q^\perp_k \to
-Q^\perp_k$ or $Q^\parallel_k \to - Q^\parallel_{-k}$.  In addition,
${\tilde V}_{kpq}$ is symmetric under permutation of $p$ and $q$, and
${\tilde V}^{(3)}_{kpq}$ is symmetric under permutation of all three
indices.  We also have ${\tilde V}_{-k,-p,-q} = - {\tilde V}_{k,p,q}$
and similarly for ${\tilde V}^{(3)}_{kpq}$.  Although the original
Hamiltonian does not have ${\tilde V}^{(3)}$ term, such a term can
present.  One of the reasons that such term can present is due to the
non-analytic behavior of the potential (because of the absolute value
$| \cdot |$).

\begin{table}
\caption{\label{tb-1}Parameters for mode-coupling equations.}
\begin{ruledtabular}
\begin{tabular}{ldddd}
Set E  &  bare  &    N=8   &   N=64    &  N=1024 \\
\hline
$c^\parallel$            &  2   &    1.435    &   1.341   &   1.329\\
$c^\perp$                &  0   &    0.621    &   0.669   &   0.674\\
$\omega^\parallel_{max}$ &  2   &    1.495    &   1.543   &   1.553\\
$\omega^\perp_{max}$     & 1.095&    1.087    &   1.182   &   1.194\\
$v_0$                    &  2   &    0.7113   &   0.6333  &   0.6337\\
$v_1$                    &  0   &   -0.1060   &  -0.0221  &  -0.0076\\
$v_2$                    & -0.6 &   -0.1099   &  -0.1242  &  -0.1293\\
$v_3$                    &  0   &    0.6362   &   0.8965  &   0.9312\\
$v_4$                    &  0   &   -0.0424   &  -0.1346  &  -0.1469\\
\end{tabular}
\end{ruledtabular}
\end{table}

We determine the parameters of the effective Hamiltonian,
$\tilde{\omega}^\parallel_k$, $\tilde{\omega}^\perp_k$, $\tilde V$,
and ${\tilde V}^{(3)}$, by fitting them to the observed
time-independent correlation functions from MD.  In order to be able
to carry out the calculation analytically, we treat the interactions
as perturbations, i.e., $\tilde V$ and ${\tilde V}^{(3)}$ are assumed
small.  All the equilibrium averages are approximated by the leading
contribution from the perturbation,
\begin{equation}
\langle \cdots \rangle \approx \langle \cdots (1 - \beta H') \rangle_0
\end{equation}
where $\langle \cdots \rangle_0$ is thermodynamical average with
respect to the non-interacting harmonic oscillators (product of
Gaussian integrals).  Under the above approximation, we find
\begin{equation}
\langle |Q^\mu_k|^2 \rangle = 
   {1 \over \beta (\tilde{\omega}^{\mu}_k)^2 } + O(\tilde{V}^2), 
\end{equation}
and the interaction parameters
\begin{eqnarray}
 \tilde{V}_{kpq} &=& { \langle Q^\parallel_k Q^\perp_p Q^\perp_q \rangle 
  \over  2 \beta \langle |Q^\parallel_k|^2 \rangle 
   \langle |Q^\perp_p|^2 \rangle \langle |Q^\perp_q|^2 \rangle }, \\
 \tilde{V}^{(3)}_{kpq} &=& { \langle Q^\parallel_k Q^\parallel_p Q^\parallel_q \rangle 
  \over  6 \beta \langle |Q^\parallel_k|^2 \rangle 
   \langle |Q^\parallel_p|^2 \rangle \langle |Q^\parallel_q|^2 \rangle }.
\end{eqnarray}
The actual form of the $\tilde \omega$ is fitted according to
Eq.~(\ref{omega_tilde_fit}).  For $\tilde V$ and ${\tilde V}^{(3)}$ we
fit into a functional form
\begin{eqnarray}
 \tilde{V}_{kpq}  =  \frac{ixyz}{\sqrt{N}} \bigl( v_0  + 
      v_1\, z^2 + v_2\, (x^2 + y^2) \bigr), \\ 
 \tilde{V}^{(3)}_{kpq} =  \frac{ixyz}{\sqrt{N}} \bigl( v_3 + 
      v_4\, (x^2 + y^2 + z^2) \bigr),
\end{eqnarray}
where $z=\sin(k\pi/N)$, $x=\sin(p\pi/N)$, and $y=\sin(q\pi/N)$.  The
values of $v_0$ to $v_4$, together with $c$ and $\omega_{max}$ are
listed in Table~\ref{tb-1}.  It is surprising that not only is the
${\tilde V}^{(3)}$ term present, but also is its magnitude comparable
to $\tilde V$.  The column marked `bare' are the parameters
corresponding to the `bare' Hamiltonian, Eq.~(\ref{H-perturb}).  The
bare parameters are reached only at very low temperatures, such as
$T=0.002$.  Since we have factored out the leading size dependence, we
expect the parameters weakly depending on size $N$.

Before presenting our results with the effective parameters, it is
worth pointing out that the standard procedure of deriving
mode-coupling equations gives identical result as our approximation,
Eq.~(\ref{our-approximation}).  The standard procedure is to project
the random force, $R_k = \tilde\omega_k^2 Q_k + \ddot Q_k$, onto a
bilinear form of the basic variable $Q_k$.  Applying the general
projection operator, Eq.~(\ref{projector}), to our case, this bilinear
projector is
\begin{equation}
 {\cal P}_2 f = 
\sum_{i\neq \bar{j}, i\leq j} \langle f Q^*_i Q^*_j \rangle 
{ Q_i Q_j \over \langle |Q_i Q_j|^2 \rangle } + 
\sum_{i} N_{i} Q_i Q_{\bar{i}},
\end{equation}
where $i$ or $j$ is a pair of indices, e.g., $j=(k,\mu)$, and $\bar
{j}$ is $(-k,\mu)$.  In evaluating the four-point correlation
functions, we have neglected the perturbation term.  It turns out that
the last term projects out only the $k=0$ component. Thus, the
specific values of $N_{i}$ are not needed.

After applying projector ${\cal P}_2$ onto $R^\mu_k$, we neglect the
difference between normal dynamics $e^{t {\cal L}}$ and anomalous
dynamics for $R^\mu_k(t)$, $e^{t {\cal P}' {\cal L}}$.  We have
\begin{eqnarray}
\Gamma^\mu_k(t) &=& \beta \langle R^\mu_k(t) R^\mu_k(0)^* \rangle
\nonumber \\
&\approx & \beta \left<
\left( e^{t{\cal L}} {\cal P}_2 R^\mu_k(0) \right)
\left( {\cal P}_2 R^\mu_k(0) \right)^* \right>.
\end{eqnarray}
The mode-coupling equations are then,
\begin{eqnarray}
\Gamma^\parallel_k(t) &=& \!\!\!\!\! \sum_{p+q=k} \!\! \left(
        \tilde{K}^\parallel_{pq} g^\perp_p(t) g^\perp_q(t)
        + \tilde{K}^{(3)}_{pq} g^\parallel_p(t) g^\parallel_q(t) \right),\\
\Gamma^\perp_k(t) &=&  \!\!\! \sum_{p+q=k} \!\!\
        \tilde{K}^\perp_{pq} g^\parallel_p(t) g^\perp_q(t),
\end{eqnarray}
where 
\begin{eqnarray}
\tilde{K}^\parallel_{pq} &=& 
\frac{\beta}{2} { |\langle R^\parallel_{p+q} Q^\perp_{-p} Q^\perp_{-q}\rangle|^2 \over 
           \langle |Q^\perp_p|^2 \rangle \langle|Q^\perp_q|^2 \rangle }  
= \frac{2}{\beta}\left| { \tilde{V}_{-p-q,p, q} \over 
              \tilde{\omega}^\perp_p \tilde{\omega}^\perp_q } \right|^2,\\
\tilde{K}^\perp_{pq} &=& 
   \beta { |\langle R^\perp_{p+q} Q^\parallel_{-p} Q^\perp_{-q}\rangle|^2 \over 
           \langle |Q^\parallel_p|^2 \rangle \langle|Q^\perp_q|^2 \rangle }  
= \frac{4}{\beta} \left| { \tilde{V}_{p, -p-q, q} \over 
              \tilde{\omega}^\parallel_p \tilde{\omega}^\perp_q } \right|^2,\\
\tilde{K}^{(3)}_{pq} &=& 
\frac{\beta}{2}{|\langle R^\parallel_{p+q} Q^\parallel_{-p} Q^\parallel_{-q}\rangle|^2 \over 
           \langle |Q^\parallel_p|^2 \rangle \langle|Q^\parallel_q|^2 \rangle }  
= \frac{18}{\beta} \left| { \tilde{V}^{(3)}_{-p-q, p, q} \over 
           \tilde{\omega}^\parallel_p \tilde{\omega}^\parallel_q } \right|^2.
\end{eqnarray}
Note that the vertex coupling $\langle R_k Q_p Q_q\rangle =
\tilde\omega^2 \langle Q_k Q_p Q_q\rangle$ is proportional to the
three-point correlation independent of the specific form of random
force.  This version of mode-coupling equation need not have the
limitation of small oscillations, and can be applied to high
temperatures as well.

\begin{figure}
\includegraphics[width=\columnwidth]{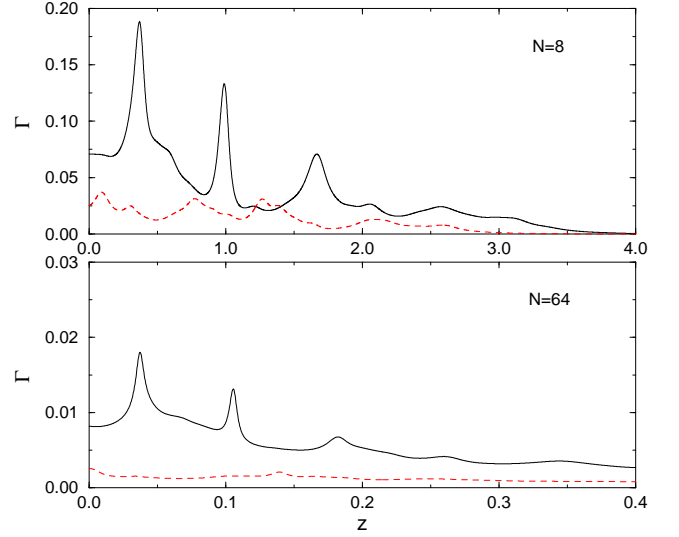}
\caption{\label{fig-MKvgm} Real part of $\Gamma_1^\parallel[z]$ (solid
line) and $\Gamma_1^\perp[z]$ (dashed line) vs $z$ from a full
mode-coupling theory for set E, using effective parameters given in
Table~\ref{tb-1}.}
\end{figure}

Figure~\ref{fig-MKvgm} is a calculation of the real part of
$\Gamma_k^\mu[z]$ for $k=1$.  Comparing with Fig.~\ref{fig-MKgm} and
\ref{fig-MKgmh}, we see that using effective parameters brings into 
much better agreement with the MD data. In fact, comparing to
Fig.~\ref{fig-MDgm}, most of the features are reproduced, such as the
locations and heights of the peaks.  The most important improvement is
the ratio of parallel to perpendicular damping.

\begin{figure}
\includegraphics[width=\columnwidth]{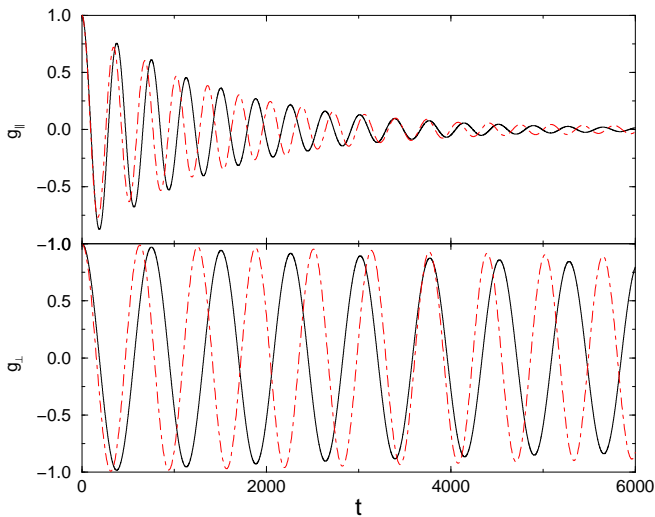}
\caption{\label{fig-MKgt}The normalized correlation functions
$g_1^\parallel(t)$ (upper part) and $g_1^\perp(t)$ (lower part) on a
$N=256$ system for data set E. The solid curves are from full
mode-coupling theory, while the dotted dashes are from equilibrium
molecular dynamics.}
\end{figure}

In Figure~\ref{fig-MKgt}, we compare the normalized dynamic
correlation functions for size $N=256$ and the slowest mode $k=1$.  The
frequencies of oscillations are slightly different in MD and
mode-coupling calculation, thus there are phase shifts at long times.
The longitudinal damping agrees with each other extremely well.  A fit
of the logarithm of amplitudes (maxima and minima) versus time for the
MD data gives the decay rate $\gamma_1^\parallel \approx 0.00064$,
$\gamma_1^\perp \approx 0.000023$, while mode-coupling predicts
$\gamma_1^\parallel \approx 0.00066$ and $\gamma_1^\perp \approx
0.000078$, respectively.  Mode-coupling theory with effective
parameters still overestimates the transverse damping by a factor of
2 to 3.

\begin{figure}
\includegraphics[width=\columnwidth]{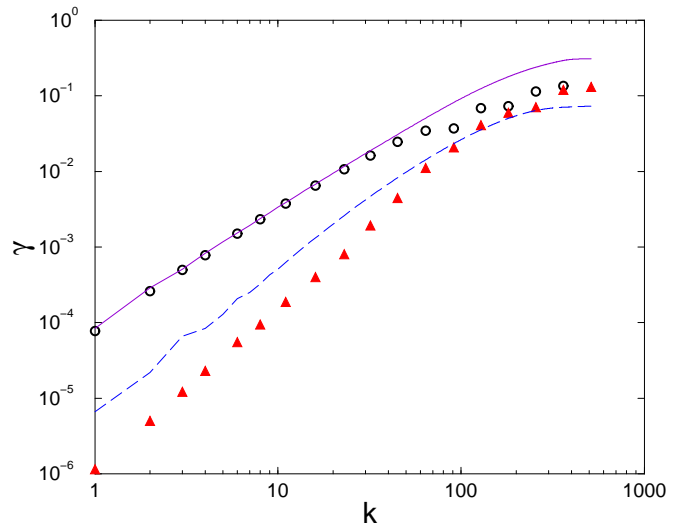}
\caption{\label{fig-gamma-compare} The decay rate $\gamma_k$ for each
mode $k$ for system size $N=1024$.  Circles ($\gamma_k^\parallel$) and
triangles ($\gamma_k^\perp$) are from molecular dynamics by fitting
the correlation function, while the continuous curves are from a full
mode-coupling theory for set E, using effective parameters given in
Table~\ref{tb-1}.}
\end{figure}

In Fig.~\ref{fig-gamma-compare}, we show the damping of the mode in
terms of the decay constant $\gamma_k$, which is defined by the
exponential decay in $g(t) \approx \cos(\tilde\omega_k t) e^{-\gamma_k
t}$.  The symbols are obtained from MD simulation, while the curves
are from the full mode-coupling theory with effective parameters.  For
MD data, $\gamma_k$ is obtained by a least-square fit to the
amplitudes.  For mode-coupling theory, we use an approximation,
$\gamma_k = \frac{1}{2} \bar \Gamma_k[\tilde \omega_k]$. The bar means
we average over a window centered around $\tilde\omega_k$ with width
about $\gamma_k$.  We observe that excellent agreement between MD data
and mode-coupling theory is obtained for the parallel component.  The
mode-coupling theory overestimates the transverse mode damping by some
constant factor.  In any case, the slopes of the curves agree very
well with the expected results (to be discussed in the next
subsection), $\gamma_k^\parallel \propto k^{3/2}$ and $\gamma_k^\perp
\propto k^2$.

\subsection{Large $N$ limit}
To simplify the equations further, we consider the large size limit.
Since the large size asymptotic behavior is the most interesting, such
limiting results are justified.  We thus consider the limit of $p \to
0$ and keep only leading contributions in lattice momentum $p= 2\pi
k/(Na)$. In this limit the three kernel functions $K_{k_1,k_2}^\mu$
become constants.  In addition, we define a $k$-independent $\nu$
function by taking the limit $p\to0$.  The leading $k$-dependence in
$\Gamma_{k}$ is $k^2$.  When this term is factored out, we expect that
$\nu_{k}$ becomes independent of $k$ in the limit of large $N$ and
small $p$.  With these simplification and approximation, and
converting the discrete summation to integral, we have in the limit of
large sizes:
\begin{eqnarray}
 \nu^\parallel(t) &=& \frac{1}{2\pi} \int_{-\pi/a}^{\pi/a}
  \!\!\!dq\, \left( K^\parallel g_q^\perp(t)^2 + 
                    K^{(3)} g_q^\parallel(t)^2\right), 
  \label{Eqsimple-mode-para}\\
 \nu^\perp(t) &=& \frac{K^\perp}{2\pi} \int_{-\pi/a}^{\pi/a}
  \!\!\!dq\, g_q^\parallel(t) g_q^\perp(t), \label{Eqsimple-mode-perp}
\end{eqnarray}
where 
\begin{equation}
K^\parallel = { a^5 K_r^2 \over 2 \beta c^\perp{}^4 m^3 }, \;
K^{(3)} = 0, \;
K^\perp = { a^5 K_r^2 \over \beta c^\parallel{}^2 c^\perp{}^2 m^3 },
\label{Kratio}
\end{equation}
if we use the perturbation expansion Hamiltonian, or
\begin{eqnarray}
K^\parallel &=& 4 \left(\frac{a}{2}\right)^7 { v_0^2 \over \beta c^\perp{}^4}, 
\quad
K^{(3)} =  
 36 \left(\frac{a}{2}\right)^7 { v_3^2 \over \beta c^\parallel{}^4}, 
\nonumber\\
K^\perp &=& 
 8 \left(\frac{a}{2}\right)^7 { v_0^2 \over 
                 \beta c^\parallel{}^2 c^\perp{}^2},
\end{eqnarray}
if we use the effective Hamiltonian.  We also linearize the dispersion
relation so that $\omega_p^\mu = c^\mu p$, $c^\mu = c^\parallel$ or
$c^\perp$ are effective sound velocities for the longitudinal and
transverse modes.

We shall refer to Eqs.~(\ref{Eqsimple-mode-para}) and
(\ref{Eqsimple-mode-perp}) as simple mode-coupling theory and the
finite $N$ version as the full mode-coupling theory.

\subsection{Asymptotic analytic solution}
Assuming that the simplified mode-coupling equations represent the
essence of the physics regarding the damping and time-dependent
correlation functions, we now consider analytic solution to
Eqs.~(\ref{Eqsimple-mode-para}) and (\ref{Eqsimple-mode-perp}).  First
we notice some constraints on the functions.  Because $\nu^\mu(t)$ is
a real function, we must have $\nu[z]^* = \nu[-z]$.  Since $g_q^\mu(0)
= 1$, we must have $\nu^\mu(0) = K^\mu/a$, for $\nu =\, \parallel$ or
$\perp$ (assuming $K^{(3)}=0$).  This implies that the Fourier-Laplace
transform must be integrable:
\begin{equation}
  \nu(0) = \frac{1}{\pi} \int_{-\infty}^\infty \! \nu[z] \,dz = \frac{K}{a}.
\end{equation}  
Next, the large $z$ behavior can be obtained by integrating by part few
times, using the boundary conditions $g(0) = 1$, $g'(0) = 0$,
$g''(0) = - \tilde \omega^2$, and $g(t\to\infty)=0$.  
For simplicity, we have omitted the
lattice momentum index $q$ and mode index $\parallel$ or $\perp$,
since it is true for any $g_q^\mu$.  With these results, we find
\begin{eqnarray}
\nu[z] &=& \int_0^\infty \!\!\nu(t)\, e^{-izt} dt  \nonumber\\
 &=& \frac{K}{2\pi} \int_{-\pi/a}^{\pi/a}\!\!\!\!\!dq \int_0^\infty\!\! 
     g_q^1(t)\, g_q^2(t)\, e^{-izt} dt \nonumber \\
&= & \frac{K}{iaz} + \frac{K}{2\pi iz^3} \int_{-\pi/a}^{\pi/a}\!\!\!\!\!\!dq(\tilde \omega_1^2 + \tilde \omega_2^2) + O(\frac{1}{z^{5}}),
\end{eqnarray}
where for $\nu^\parallel$, $g^1$ and $g^2$ are all $g^\perp$, while
for $\nu^\perp$, it is $g^\parallel$ and $g^\perp$.

We can establish that
\begin{equation}
\nu^\parallel[z] \propto (iz)^{-1/2}, \quad \nu^\perp[z] \propto {\rm const},
\end{equation}
in the limit of small $K^\perp$.  We proceed as follows: first we
assume that $\nu^\perp[z] = \nu_0^\perp$ or $\nu^\perp(t) =
2\nu_0^\perp \delta(t)$.  We then derive an expression for
$\nu^\parallel(t)$ and $\nu^\perp(t)$, and show that indeed
$\nu^\perp(t)$ approaches a $\delta$-function in a proper limit.

When $\nu^\perp[z]$ is a constant, the inverse Fourier transform for
the correlation function $g[z]$ can be performed exactly, to give the
transverse correlation function as
\begin{eqnarray}
g_q^\perp(t) &=& \frac{1}{2} \left(1 + \frac{\Gamma_q^\perp}{2i \bar\omega}
\right) e^{i\bar\omega t - \frac{1}{2}\Gamma_q^\perp t} +  \nonumber\\ 
&&  \frac{1}{2} \left(1 - \frac{\Gamma_q^\perp}{2i \bar\omega}
\right) e^{-i\bar\omega t - \frac{1}{2}\Gamma_q^\perp t}, 
\end{eqnarray}
where $\Gamma_q^\perp = \nu_0^\perp q^2$, and $\bar\omega =
\sqrt{c^\perp{}^2 q^2 - \Gamma_q^\perp{}^2/4} \approx c^\perp q$.  We
have linearized the dispersion relation, $\tilde \omega \approx
c^\perp q$.  We only look at the dominant term in time dependence of
$\nu^\parallel(t)$.  For $g_q^\perp(t)^2$, since large $q$-mode decays
rapidly, we ignore the term with amplitude proportional to $q^2$, and
also drop the oscillatory term and approximate, $g_q^\perp(t)^2
\approx \frac{1}{2} e^{-\Gamma_q^\perp t}$.  After integrating over $q$ (we also
extend the limit as from $-\infty$ to $\infty$), we obtain the leading
$t$ dependence, as
\begin{equation}
\nu^\parallel(t) \approx 
\frac{K^\parallel}{4} \sqrt{ \frac{1}{\pi \nu_0^\perp t}}.
\end{equation}
If the oscillatory terms are kept, they only contribute an
exponentially small term, $e^{-(c^\perp)^2t/\nu_0^\perp}$; and the $q^2$ terms
give a contribution that decays much faster, as $t^{-3/2}$.  The
contribution from $K^{(3)}$ term can be neglected, because it decays
as $t^{-2/3}$ (due to Eq.~(\ref{gpara-of-t})).  But this term does
provide a crossover from $z^{-1/3}$ for intermediate $z$ to the
asymptotic $z^{-1/2}$ for very small $z$.  The Fourier-Laplace
transform gives
\begin{equation}
\nu^\parallel[z] \approx \frac{K^\parallel}{4}  
\sqrt{ \frac{1}{i\nu_0^\perp z}} = b (iz)^{-1/2}.
\label{nu-para-z}
\end{equation}

We now get an expression for the longitudinal correlation function.
Formally, it is given by the inverse transform
\begin{equation}
g_q^\parallel(t) = 
\frac{1}{2\pi} \int_{-\infty}^\infty 
{-iz - \Gamma_q^\parallel[z] \over
 z^2 - c^\parallel{}^2 q^2 - i z \Gamma_q^\parallel[z] } e^{izt} dz,
\end{equation}
where $\Gamma_q^\parallel[z] = q^2 \nu^\parallel[z] = bq^2/\sqrt{iz}$,
$b= \frac{1}{4}K^\parallel/\sqrt{\nu_0^\perp}$.  The dominant
contribution is from $z$ when the denominator is close to zero.  The
integral can be approximately estimated by the residue theorem.  By
location the poles
\begin{equation}
 z^2 -c^\parallel{}^2 q^2 - \sqrt{iz}\, b\, q^2 = 0, 
\end{equation}
we obtain the dispersion relation for $g_q^\parallel(t)$.  In the limit of small
$q$, we find
\begin{equation}
z \approx c^\parallel q + i \gamma_0 q^{3/2},
\end{equation}
where $\gamma_0 = \frac{\sqrt{2}}{16} K^\parallel / \sqrt{c^\parallel
\nu_0^\perp}$.  Therefore,
\begin{equation}
 g_q^\parallel(t) \approx e^{-\gamma_0 q^{3/2} t} \cos(c^\parallel qt).
\label{gpara-of-t}
\end{equation}
Similarly with the dispersion relation for the transverse mode, $z
\approx c^\perp q + \frac{1}{2} i \nu_0 q^2$, we can compute
\begin{eqnarray}
\nu^\perp(t) &=& \frac{K^\perp}{\pi} \int_{0}^\infty\!\!\! dq\,
e^{-\gamma_0 q^{3/2}t - \frac{1}{2} \nu_0^\perp q^2 t} \cos(c^\parallel qt)
\cos(c^\perp qt) \nonumber \\
& \approx & K^\perp \sqrt{\frac{1}{8\pi\nu_0^\perp t}} \Bigl( 
e^{-\frac{(c^\parallel - c^\perp)^2}{2\nu_0^\perp} t} + 
e^{-\frac{(c^\parallel + c^\perp)^2}{2\nu_0^\perp} t} \Bigr). 
\end{eqnarray}
We have neglected the $\gamma_0$ term.  Since $|c^\parallel -
c^\perp|$ is smaller than $c^\parallel + c^\perp$, we drop the second
term.  We symmetrically extend the function.  Note that $\nu^\perp(t)$
can be casted into the functional form
\begin{equation}
f_\sigma(t) = \frac{1}{2} \sqrt{\frac{\sigma}{\pi}} { e^{-\sigma|t|} \over \sqrt{|t|} },
\quad \sigma = { (c^\parallel - c^\perp)^2 \over 2 \nu_0^\perp },
\end{equation}
which has the property that
\begin{equation}
\int_{-\infty}^\infty f_\sigma(t)\, dt = 1, \quad {\rm for\ all}\quad \sigma > 0.
\end{equation}
Thus $f_\sigma(t)$ behaves as a $\delta$-function as $\sigma \to
\infty$:
\begin{equation}
\nu^\perp(t) \approx { K^\perp \over | c^\parallel - c^\perp| } \delta(t).
\end{equation}
This allows us to identify the constant $\nu_0^\perp = \frac{1}{2}
K^\perp/|c^\parallel - c^\perp|$.  The condition for $\sigma \to
\infty$ is the same as $\nu_0^\perp \to 0$ or $K^\perp \to 0$.  Since
$K^\perp/K^\parallel = 2 (c^\perp/c^\parallel)^2$
(c.f. Eq.(\ref{Kratio})), small $K^\perp$ also implies small
$c^\perp$.

Although the above derivation suggests that the result is valid only
for the special limit.  But in fact, this is also the asymptotic
result in the limit $z \to 0$.  To show this, we note that the
asymptotic behavior is picked up by the scaling $\lim_{\lambda \to
0}\lambda^{\delta} \nu[\lambda z] \approx \nu[z]$.  If this is true,
then $\nu[z] \propto z^{-\delta}$.  This scaling limit coincides with
the limit of $K^\perp \to 0$.

\subsection{Numerical solution of the simple theory}

For the simple mode-coupling theory that we have already taken the
large $N$ limit, we can also apply the fast Fourier transform method.
However, we find that a direct numerical integration in frequency
space is much more accurate.  For example, $\nu^\perp[z]$ is expressed
as
\begin{eqnarray}
\nu^\perp[z] &= & { K^\perp \over (2\pi)^3 } 
\int_{-\infty}^{+\infty}\!\!\!\!\!\!\! d\omega 
\int_{-\infty}^{+\infty}\!\!\!\!\!\!\! d\omega'
\,2\,F_\perp(\omega, \omega') \nonumber \\
& & \left( {\rm P} { 1 \over i(z-\omega-\omega') } 
+ \pi \delta(z-\omega-\omega') \right), 
\label{principle-value-int}
\end{eqnarray}
where P stands for Cauchy principal value and 
\begin{eqnarray}
F^\perp(\omega, \omega') &=& \int_0^{\pi/a} \!\!g^{\parallel}(q,\omega) 
g^{\perp}(q,\omega')\, dq,\\
g^{\parallel,\perp}(q,\omega) &=& { - i \omega - \nu^{\parallel,\perp}[\omega] q^2 \over 
          \omega^2 - (c^{\parallel,\perp})^2 q^2 - i\omega \nu^{\parallel,\perp}[\omega] q^2 }.
\end{eqnarray}
When the linear approximation is made to the dispersion relations, the
integral over $q$ can be performed analytically.  We only need to do a
two-dimensional integral over $\omega$ and $\omega'$.  The principal
value integral is taken care by locating exactly the singularity.
Since the integration routine needs a smooth function $\nu[z]$, we fit
the results by a Pad\'e approximation (in variable $iz$ or
$(iz)^{1/3}$).  The procedure is iterated several times for
convergence.  This is programmed in \textsl{Mathematica}.  It turns
out essential to have more than double precision accuracy in the
integration routine in order for the singular integrals properly
converged.  Some results of this calculation are already presented in
ref.~\cite{Wang-Li}.

\begin{figure}
\includegraphics[width=\columnwidth]{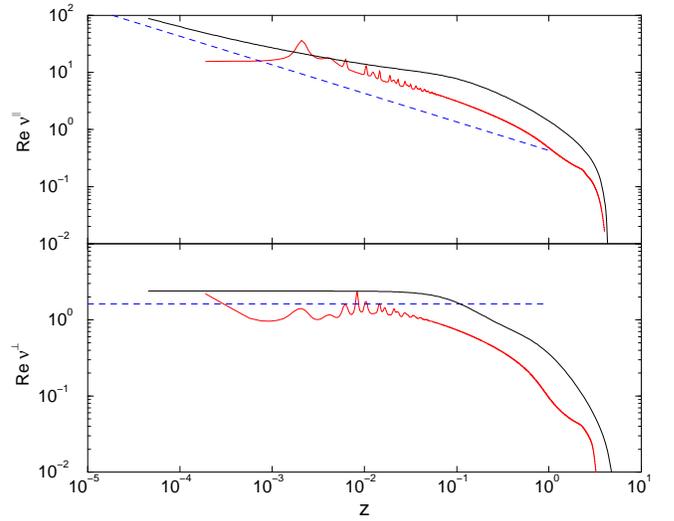}
\caption{\label{fig-nu} Real part of $\nu^\parallel [z]$ (top) and 
$\nu^\perp[z]$ (bottom) for data set E.  The smooth curves are from
simple mode-coupling theory, the (red) curves with spikes are computed
from full theory at $N=1024$, while the straight, dash lines are
analytic asymptotic results.}
\end{figure}
\begin{figure}
\includegraphics[width=\columnwidth]{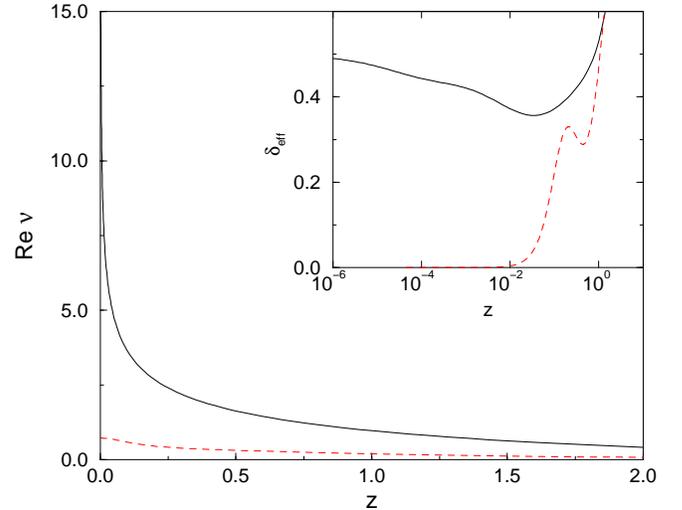}
\caption{\label{fig-nu-crossover} Real part of $\nu^\parallel [z]$ 
(solid line) and $\nu^\perp[z]$ (dashed line) for data set J
($K^\parallel=1.532$, $K^\perp=0.719$, and $K^{(3)}=2.776$). The
insert shows the effective exponent $-d \ln \nu/d \ln z$. }
\end{figure}

In Fig.~\ref{fig-nu}, we compare three levels of approximations of
$\nu^\mu[z]$: the simple mode-coupling theory, the full theory
computed on $N=1024$ by $\nu^\mu[z] = \Gamma_1^\mu[z]/(2\pi/(Na))^2$,
and the asymptotic result of $\nu^\perp[z] = \frac{1}{2} K^\perp
\bigl(1/|c^\parallel - c^\perp| + 1/|c^\parallel + c^\perp|\bigr)$,
and Eq.~(\ref{nu-para-z}).  Noticeable differences are seen between
full theory and simplified.  This is partly due to the fact that we
are comparing a finite $N$ with an infinite $N$ result.  We also
note that the asymptotic slope of $-1/2$ for $\nu^\parallel[z]$ is
approached rather slowly.

Figure~\ref{fig-nu-crossover} shows a stronger crossover effect from
slope of $1/3$ (corresponding to $\kappa \propto N^{2/5}$, to be
discussed later) to that of the true asymptotic law of 1/2.  This set
of curves corresponds to data set J.

\subsection{Scaling solution of the simple theory}
\begin{figure}
\includegraphics[width=\columnwidth]{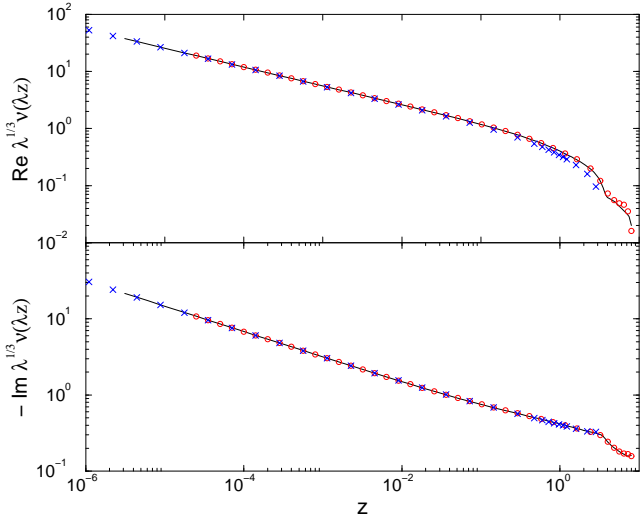}
\caption{\label{fig-lepri-scaling} Real part (top) and imaginary part
(bottom) of $\lambda^{1/3} \nu(\lambda z; K, \lambda^{1/3}c,
\lambda^{-2/3} a)$ vs $z$ for $K=c=a=1$ and $\lambda =1 $ (solid
line), 1/8 (open circles), and 8 (crosses).  }
\end{figure}

We now look at some of the scaling properties that are implied from
Eqs.~(\ref{Eqsimple-mode-para}) and (\ref{Eqsimple-mode-perp}).
First, we consider the simple case of $K=K^\parallel =K^\perp$,
$K^{(3)}=0$, and $c=c^\parallel = c^\perp$.  In this case the two
equations degenerate into one equation given by Lepri
\cite{lepri-mode-coupling}.  By measuring frequency in terms of $c$
(i.e., consider variable $z/c$) we can scale away $c$, thus the
following equation is an exact scaling
\begin{equation}
\nu(z;K,c,a) = c\, \nu(z/c; K/c^2, 1, a),
\end{equation}
where $\nu$ is a function of $z$ with parameters $K$, $c$, and $a$
that we have written out explicitly.  This equation tells us that out
of the three parameters that characterize the solution, we can pick
two as independent.  The ``shapes'' of the solutions are all the same
for $K/c^2 = {\rm const}$. Without loss of generality, we can fix
$K=1$.

Next, we consider a general scaling solution of the form
\begin{equation}
\nu(\lambda z; K, \lambda^{\Delta_c} c, \lambda^{\Delta_a} a)  
\approx \lambda^{-\delta} \nu(z;K,c,a).
\label{lepri-scaling}
\end{equation}
Substituting this scaling form into Eq.~(\ref{principle-value-int}),
by requiring that the result must be consistent, we obtain the
following conditions
\begin{eqnarray}
2 \Delta_q - \delta - 1 & = & 0,\\
\Delta_c + \Delta_q - 1 & = & 0,\\
1 - \Delta_q & = & \delta, \\
\Delta_a + \Delta_q &=& 0, 
\end{eqnarray}
where $\Delta_q$ is scaling exponent associated with integration
variable $q$, $q \to \lambda^{\Delta_q} q$.  A unique solution to the
set of linear equations is obtained, $\delta = 1/3$, $\Delta_c = 1/3$,
$\Delta_a = - 2/3$, and $\Delta_q = 2/3$.  Since
Eq.~(\ref{lepri-scaling}) is (approximately) valid for any $\lambda$,
we can choose $\lambda = 1/z$.  This scaling solution implies that
\begin{equation}
\nu(z;K,c,a) \approx z^{-1/3} \nu(1; K, c/z^{1/3}, z^{2/3}a).
\label{lepri-scaling-numeric}
\end{equation}
Power-law behavior $z^{-1/3}$ is obtained in the limit of small $z$,
relatively large $c$, and small $a$, if $\nu(1, K, \infty, 0)$ is
finite.  The crossover to other behavior occurs at large $z \sim c^3$
and $z \sim a^{-3/2}$.

A simple dimension analysis also leads to the $z^{-1/3}$ factor.  Let
the dimension of length and time be $L$ and $T$, respectively.  Then
the dimensions of relevant quantities are $[z] = T^{-1}$, $[c] =
LT^{-1}$, $[a] = L$, $[\nu] = L^2 T^{-1}$, and $[K] = L^3 T^{-2}$.
From the five quantities, we can construct three dimensionless
variables
\begin{equation}
\Pi_1 = { \nu \over z^{-1/3} K^{2/3} }, \quad
\Pi_2 = { c \over z\,a }, \quad
\Pi_3 = { K \over a c^2 }.
\end{equation}
If there is any relation between the $\Pi_i$'s, it must be in the form
$\Pi_1 = f(\Pi_2, \Pi_3)$ (Buckingham Pi theorem \cite{buckingham}),
or
\begin{equation}
 \nu[z] = z^{-1/3} K^{2/3} f\left( { z a\over c}, {K \over a c^2 }\right), 
\end{equation}
where $f$ is an arbitrary, dimensionless function.  This result, of
course, is consistent with Eq.~(\ref{lepri-scaling}).  This also
suggests that the scaling is not approximate, but an exact result.

Fig.~\ref{fig-lepri-scaling} is a test of the above scaling by
numerical solutions.  For small $z$, the power-law $z^{-1/3}$ is
verified to high accuracy.  Since this scaling is exact, the
deviations are due purely to numerical errors in solving the
equations.

The case of the coupled longitudinal and transverse equations is
somewhat difficult to analyze.  We can require a very general scaling
of the form
\begin{eqnarray}
\nu^\parallel(\lambda z; \lambda^{\Delta_{K^\parallel}} K^\parallel, 
\lambda^{\Delta_{K^\perp}} K^\perp, 
\lambda^{\Delta_{c^\parallel}} c^\parallel, 
\lambda^{\Delta_{c^\perp}} c^\perp, 
\lambda^{\Delta_a} a) &&\nonumber \\ 
\approx \lambda^{-\delta_\parallel} 
\nu(z;K^\parallel,K^\perp,c^\parallel,c^\perp,a),\quad &&
\end{eqnarray}
and similarly for the perpendicular component.  If we require for
consistency of scaling for both longitudinal and transverse
components, we find that the only scaling solution is the symmetric
solution, i.e., the scaling discussed above with identical
longitudinal and transverse scaling exponents.

If we abandon the exact scaling for the transverse component and look
only at longitudinal component, we can require that
\begin{eqnarray}
\delta_\perp &=& 2 \Delta_q - 1,\\
\delta_\parallel &=& 1 - \Delta_{K^\parallel} - \Delta_q, \\
\Delta_{c^\perp} &=& 1  - \Delta_q.
\end{eqnarray}
As before we can require that there is no scaling for the coupling
constant $K^\parallel$ without loss of generality (i.e., 
$\Delta_{K^\parallel}=0$), the above equations imply a relation
\begin{equation}
2 \delta_\parallel + \delta_\perp = 1.
\end{equation}
In particular, if $\delta_\perp = 0$, we must have $\delta_\parallel =
1/2$ and $\Delta_{c^\perp} = 1/2$.  This is consistent with the fact
that the scaling region of $\delta_\parallel = 1/2$ is obtained for
small $c^\perp$.

A complete and clear scaling picture is still lacking.  From the
numerical solution, we observe that besides well characterized scaling
regimes, there are also plenty of intermediate regions and crossovers.
For very large $c^\perp$ and small $K^\perp$, or small $c^\perp$ and
large $K^\perp$, the behavior of the solutions are difficult to
characterize.

\section{\label{sec:green-kubo}Green-Kubo Formula}
In this section, we make the connection of the damping of modes with
thermal conduction.  The starting point is the Green-Kubo formula for
transport coefficient:
\begin{equation}
 \kappa = \frac{1}{k_B T^2 a N} \int_0^\infty \! \langle J(t) J(0) \rangle\,dt.
\end{equation}
For the special case of heat conduction in one-dimensional chain, it
is better to call $\kappa$ heat conductance rather than heat
conductivity.  In analogous to electric circuit, the heat conductance
relates the temperature gradient to energy current (Fourier law),
\begin{equation}
 I = - \kappa { dT \over dx}.
\end{equation}
The quantity $J(t)$ is related to energy current by $J = I aN$.

The central quantity that we compute in equilibrium and nonequilibrium
heat conduction is the (total) heat current $J$.  Since the heat
current is a macroscopic concept determined by the conservation of
(internal) energy, a microscopic version of it is not unique.  The
current expression must satisfy the energy continuity equation in the
long-wave limit.  We derive an expression starting from ${\bf J} =
\sum_i d( {\bf r}_i h_i) /dt$, where the local energy per particle is
$h_i = \frac{1}{4} K_r \bigl( ( |\Delta {\bf r}_{i-1}|-a)^2 + (|\Delta
{\bf r}_{i}| -a)^2\bigr) + K_\phi \cos \phi_i + p_i^2/(2m)$.  By
regrouping some of the terms using translational invariance, we arrive
at the heat current per particle:
\begin{eqnarray}
m\, {\bf j}_i &=&  -\, \Delta {\bf r}_i \bigl(( {\bf p}_i + 
{\bf p}_{i+1} ) \cdot {\bf G}(i) \bigr)
\nonumber\\ 
&& -\, \Delta {\bf r}_{i-1} \bigl( ( {\bf p}_i + {\bf p}_{i-1} ) 
\cdot {\bf G}(i-1) \bigr)
\nonumber\\
&& +\, \Delta {\bf r}_{i-1} \bigl( {\bf p}_i \cdot {\bf H}(i\!-\!2,i\!-\!1,i\!-\!1)\bigr)
\nonumber\\
&& +\, \Delta {\bf r}_{i} \bigl( {\bf p}_i \cdot {\bf H}(i\!+\!1,i\!+\!1,i) \bigr)
+ {\bf p}_i h_i, 
\end{eqnarray}
where ${\bf G}(i) = \frac{1}{4} K_r \bigl(|\Delta {\bf r}_i | -
a\bigr) {\bf n}_i$, ${\bf H}(i,j,k) = K_\phi( {\bf n}_i + {\bf n}_k
\cos \phi_j) / | \Delta {\bf r}_k |$, ${\bf n}_i$ is a unit vector in
direction of $\Delta {\bf r}_i = {\bf r}_{i+1} - {\bf r}_i$.  The
total heat current in $x$ direction, $J = \sum_i j_{x,i}$ is the
quantity appearing in the Green-Kubo formula.  It is equal to the
macroscopic heat current density (energy per unit area per unit time)
integrated over a volume.

For theoretical analysis, we need to express $J$ in terms of the
dynamical variables $P_k$ and $Q_k$.  A general expression will be
rather complicated. Again as in mode-coupling approach, we'll consider
small oscillation expansions, and consider only the leading
contributions.  Neglecting the nonlinear contribution of $O(Q_k^3)$,
we obtain
\begin{equation}
J = \sum_{k,\mu} b_k^\mu Q_k^\mu P_{-k}^\mu,
\quad b_k^\mu = i\, \omega_k^\mu { \partial \omega_k^\mu \over 
                           \partial \left({ 2\pi k \over Na}\right)  },
\end{equation}
where $\omega_k^{\mu}$ is the bare dispersion relation given by
Eqs.~(\ref{bare-1}) and (\ref{bare-2}).  More general expression in
a quantum-mechanical framework including the cubic terms is given in
ref.~\cite{hardy}.

An approximation for the correlation function of the current can be
obtained, again using a dynamic mean-field approximation, as
\begin{eqnarray}
\langle J(t) J(0) \rangle & = & 
\sum_{k,\mu} |b_k^\mu|^2 \bigl\{
\langle Q_k^\mu(t) Q_{-k}^\mu(0) \rangle
\langle P_k^\mu(t) P_{-k}^\mu(0) \rangle \nonumber \\
 & & + \langle Q_k^\mu(t) P_{-k}^\mu(0) \rangle^2 \bigr\}.
\label{GK-mkp}
\end{eqnarray}

The above expression can be further simplified by the approximation
$P_k^\mu = \dot Q_k^\mu \approx \tilde\omega_k^\mu Q_k^\mu$.  We find
\begin{eqnarray}
\langle J(t) J(0) \rangle & \approx & 
\sum_{k,\mu} \left| { b_k^\mu \over \beta \tilde \omega_k^\mu} 
            \right|^2 g_k^\mu(t)^2 \nonumber \\
 & \approx &  { aN (c^\parallel)^2 \over 2 \pi \beta^2} 
              \int_0^\infty\!\! \cos^2(\tilde\omega_p^\parallel t)\, 
              e^{-2\gamma_p^\parallel t} dp  \nonumber \\
 & \propto & 
 { N (c^\parallel)^2 \over \beta^2 } t^{-1/(2-\delta_\parallel)}.
\label{mk-t-dependence}
\end{eqnarray}
We have dropped the contribution from the perpendicular component,
because it decays much faster (due to an extra $p^4$ factor in the
integrand).

\begin{figure}
\includegraphics[width=\columnwidth]{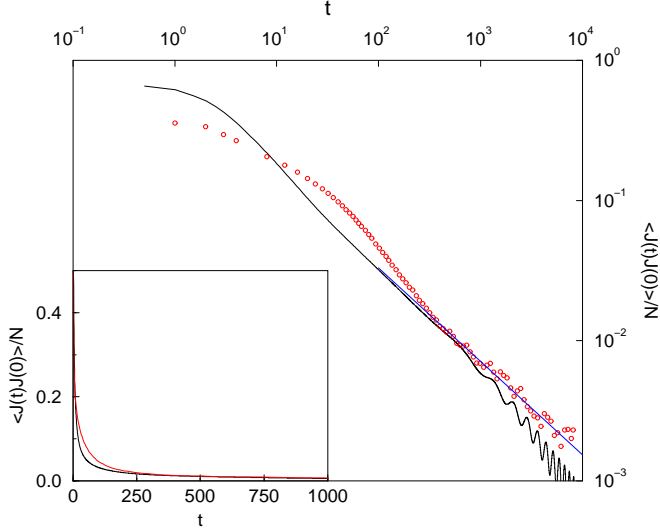}
\caption{\label{fig-Jt-compare} The Green-Kubo integrand from MD
(circles) and mode-coupling theory (curve) for set E with size
$N=1024$.  The straight has a slope $-2/3$.  The insert shows the same
data on a linear scale.  }
\end{figure}

We compare the mode-coupling result of Green-Kubo integrand with that
of MD in Fig.~\ref{fig-Jt-compare}.  We have used Eq.~(\ref{GK-mkp})
for the mode-coupling calculation.  Note that the correlation
functions $\langle Q(t) Q(0)^* \rangle$, $\langle P(t) P(0)^*
\rangle$, and $\langle Q(t) P(0)^* \rangle$ are simply related in
frequency domain through Eq.~(\ref{general-Gz}).  The agreement is
reasonable with the largest deviation about a factor of 2.  The asymptotic
behavior of $t^{-2/3}$ is consistent with both sets of results.

\begin{figure}
\includegraphics[width=\columnwidth]{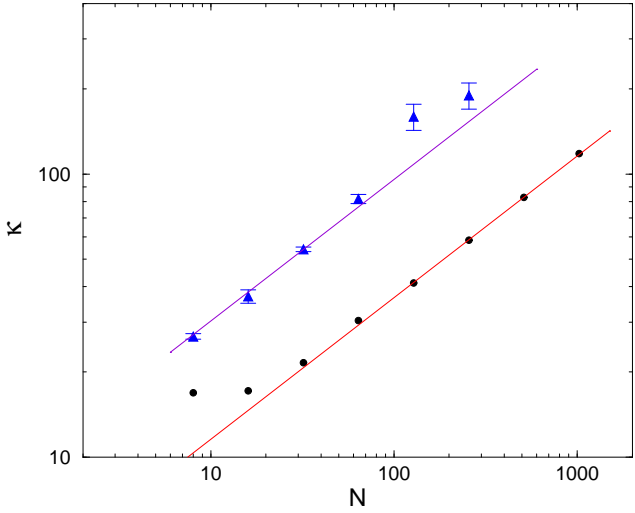}
\caption{\label{fig-kappa}The finite lattice conductance 
defined by Green-Kubo formula at parameter set E.  Solid circles are 
from full mode-coupling theory; the triangles are from
molecular dynamics with periodic boundary condition; the straight
lines have slope of 1/2.} 
\end{figure}

The heat conductance on a finite lattice is sensitive to boundary
conditions. Clearly, in the mode-coupling formulation, we have used
periodic boundary condition.  If we integrate over time from 0 to
$\infty$ first on the second line of Eq.~(\ref{mk-t-dependence}), we
obtain $\kappa \propto \int dp/\gamma_p$. On a finite lattice we have
a lower momentum cut-off, $2\pi/(Na)$.  The size dependence of the
conductance on finite lattice is then
\begin{equation}
 \kappa_N \propto N^{1-\delta_\parallel}.
\end{equation}
Since $\delta_\parallel = 1/2$, $\kappa \propto \sqrt{N}$.  This
result is in agreement with that of Deutsch and Narayan for a model
(the random collision model) where transverse motion is taken into
account only stochastically \cite{deutsch-narayan}.

In Fig.~\ref{fig-kappa}, the mode-coupling result of $\kappa_N$ is
compared with equilibrium molecular dynamics result with periodic
boundary conditions.  Excellent 1/2 power is observed for the
mode-coupling result.  We found it is rather difficult to get
converged value from molecular dynamics.  But the results are
consistent with a $\sqrt{N}$ law.

The relation of the mode-coupling theory with the results of
nonequilibrium situation of low and high temperature heat-baths at the
ends is not clear cut.  The standard assumption is that the
correlation should be cut off by a time scale of order $N$ due to the
interaction with the heat baths.  Finite size result is obtained by
integrating the power-law decay to a time of $O(N)$.  This gives us
the asymptotic behavior for the conductance at large $N$ as
\begin{equation}
\kappa_N \propto N^{\alpha}, \quad
\alpha = 1 - { 1 \over 2 -\delta_\parallel }.
\end{equation}
Since we find $\delta_\parallel = 1/2$ for small $K^\perp$, the
thermal conduction diverges as $N^{1/3}$.  When $K^{(3)}$ is
sufficiently large, we observe $N^{2/5}$.

\section{\label{sec:MD}Nonequilibrium Molecular Dynamics Study}

\subsection{MD simulation details}
 
We note that the interaction potential is not smooth at the point when
two particles overlap. This can cause numerical instability.  Thus, we
replaced the original potential with a modified one,
\begin{equation}
\Delta r = | {\bf r}_{i+1} - {\bf r}_i| \quad \to\quad
\Delta r' = \Delta r + {\epsilon^2 \over \Delta r + \epsilon }, 
\end{equation}
which smooths out the discontinuous derivatives at $\Delta r=0$ with a
small correction of order $\epsilon^2$.  In addition to replacing the
spring potential by $\frac{1}{2} K_r (\Delta r' - a)^2$, we also
replace the cosine term by
\begin{equation}
\cos \phi'_i = - { \Delta {\bf r}_{i-1} \cdot \Delta {\bf r}_i \over
                   \Delta r'_{i-1} \Delta r'_{i} },
\end{equation}
so that the value is well-defined for all positions ${\bf r}_i$. 

In actual simulation, we have used very small $\epsilon \sim 10^{-3}$
to $10^{-4}$ so that its effect should be comparable to error caused
by finite time-step $h$.  We also used large $\epsilon$ as a way of
simulating a slightly different model to study the robustness of the
results.

We solve the system of equations \cite{frenkel-smit}
\begin{equation}
 { d{\bf p}_i \over dt} = 
\left\{
 \begin{array}{ll}
 {\bf f}_i - \xi_L {\bf p}_i, & \mbox{if $ i < N_w$;}\\
        {\bf f}_i, & \mbox{if $ N-N_w > i \geq  N_w$;}\\
        {\bf f}_i - \xi_H {\bf p}_i, & \mbox{if $ i \geq N-N_w$;}
 \end{array}
\right.
\end{equation}
where ${\bf f}_i$ is the total force acting on the $i$-th particle,
$\xi_L$ and $\xi_H$ obey the equation
\begin{equation}
 {d \xi_{L,H} \over dt} =  
{ 1 \over \Theta^2} \left( - 1 +   { 1 \over k_B T_{L,H} N_w }
\sum_{i} { {\bf p}_i^2 \over m}  \right), 
\end{equation}
where $T_L$ and $T_H$ are the temperatures of the two heat baths at
the ends. The summation is over the particles belonging to the heat
bath.  We have used 4 particles for the Nos\'e-Hoover heat baths, with
extra first two and last two particles at fixed positions.  The
coupling parameter $\Theta$ is taking to be 1.  For the central part,
we used a second-order symplectic algorithm (or equivalently, the
velocity Verlet algorithm), while for the heat-bath, we used a simple
difference scheme accurate to second order in $h$ for $\xi$.

Although the total energy (Hamiltonian) is no longer a conserved
quantity when the heat-bath is introduced, the above equation still
has a conserved quantity of similar character:
\begin{equation}
H(p,q) + \sum_{x=L,H} N_w k_B T_x \left(
\frac{1}{2}\Theta^2 \xi_x^2 + \int_0^t \xi_x dt \right).
\label{conserved}
\end{equation}
This quantity can be used to monitor the stability of the algorithm.

Since we run for very long time steps of $10^8$ to $10^{10}$, it is
essential that the algorithm is stable over extended period of time.
Even with a symplectic algorithm, stability is not guaranteed by
merely taking small $h$ $(\sim 10^{-4})$.  Thus, we adjusted 
the conserved quantity, Eq.~(\ref{conserved}), to
its starting value after certain number of steps.

\subsection{Heat conductance results}

\begin{figure}
\includegraphics[width=\columnwidth]{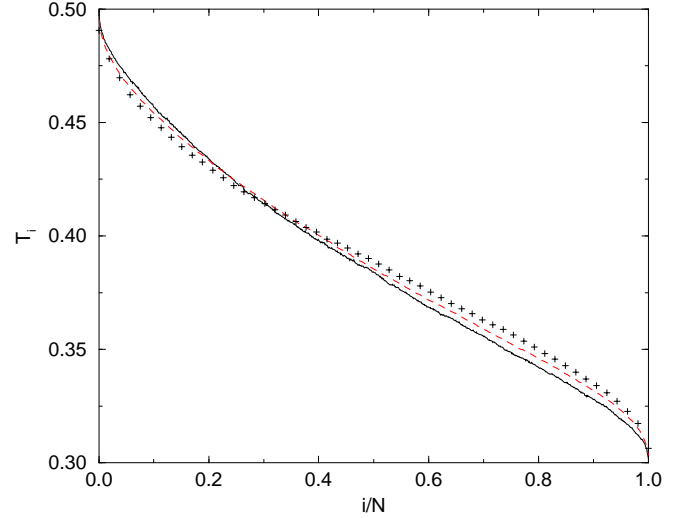}
\caption{\label{fig-temperature-scaling}Temperature of $i$-th
particle, $T_i$ vs scaled position, $i/N$, for data set E with $N= 64$
(plus), 256 (dash line), and 1024 (solid line).  }
\end{figure}

In Fig.~\ref{fig-temperature-scaling}, we observe that good
temperature profiles are established.  Due to relatively large
temperature difference between low and high temperature heat baths,
and perhaps also due to the nature of heat baths, the profile is not
linear.  However, the scaling with $N$ is approximately obeyed.

\begin{figure}
\includegraphics[width=\columnwidth]{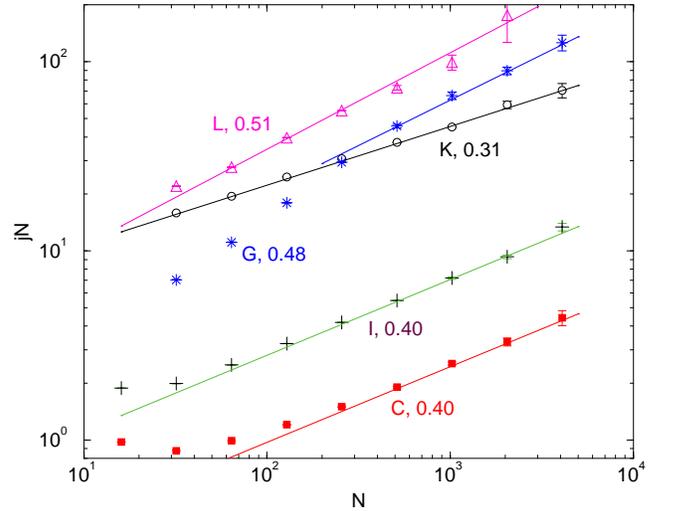}
\caption[long]{\label{fig-jL-extra}$j N$ vs $N$ for different
parameters $(K_\phi, T_L, T_H, \epsilon)$.  set C: $(K_\phi, T_L, T_H,
\epsilon) = (0.1, 0.2, 0.4, 0)$, the chain is compressed to have an
average distance between particle $d_0=1.5$; set G: (10, 0.2, 0.4, 0);
set I: (0.1, 0.3, 0.5, 0.2); set K: (0.5, 1.2, 2, 0.4); set L: (25, 1,
1.5, 0.2).  All of the sets have $K_r=1$, mass $m=1$, and lattice
constant $a=2$.  The number indicates the slope of least-squares fit.
}
\end{figure}

In addition to the thermal conductance results reported in
ref.~\cite{Wang-Li}, Fig.~\ref{fig-jL-extra} gives additional data for
variety of parameters.  One of the aim of these additional runs is to
check the robustness of the $1/3$ power law for thermal conductance.
It is found in ref~\cite{Wang-Li} for set E exponent $\alpha = 0.334
\pm 0.003$.  This is an excellent confirmation of the mode-coupling
theory.  We note that at parameter set $K$, the $1/3$ law is not
destroyed by introducing large $\epsilon$.  Thus we believe that the
$1/3$ law is not specific to the potential used in
ref.~\cite{Wang-Li}.  In fact, the Fermi-Pasta-Ulam model with transverse
motion was studied
\cite{Li-Han-Wang} and result is also consistent with the $1/3$ power
law.

At low temperatures for set I and C, we appear to observe exponent of
0.4.  What is particularly interesting for set C is that the chain is
compressed to an average distance of 1.5 from the equilibrium distance
of 2.  The behavior is not changed much comparing to uncompressed
chain.  Set G and L represent very large angular coupling $K_\phi$.
We note that $K_\phi \to \infty$ corresponds to a situation that
equilibrium can not be established. As $K_\phi$ becomes larger, the
equilibration becomes increasingly difficult.  In ref.~\cite{Wang-Li}
we reported a logarithmic behavior for $K_\phi = 1$ (set B).  This
logarithmic behavior does not maintain when $K_\phi$ is increased
further. The exponent $\alpha$ for set L and G is closed to 1/2.  This
may be related the results of disordered harmonic chain.

\section{Conclusion}

Lepri {\sl et al} first introduced mode-coupling theory into the
problem of heat conduction in low dimensional systems to interpret the
MD results qualitatively.  We have developed the theory further by
introducing a `full theory' and effective parameters for theory.  The
full theory is solved numerically.  At this level of approximation, we
find that mode-coupling theory gives excellent prediction on a
quantitative level.  The longitudinal damping (decay rate)
$\gamma_k^\parallel$ agrees with MD data with few per cents of
deviation.  The mode-coupling theory somewhat overestimated the
transverse damping $\gamma_k^\perp$ by a factor of two to three.  The
full theory assumes that the damping function (memory kernel)
$\Gamma_k[z]$ is an explicit function of two variables, $k$ and $z$.
In a simple mode-coupling theory, we assume $\Gamma_k[z] \approx (2\pi
k/(Na))^2 \nu[z]$, valid for small $k/N$.  The simple theory is
computationally efficient, and contains essential asymptotic features,
i.e., $\gamma_k^\parallel \propto k^{3/2}$ and $\gamma_k^\perp \propto
k^2$.  Through detailed comparison between MD and mode-coupling theory
on memory kernel, decay of the modes, the time-dependent correlation
functions, Green-Kubo integrand, and finally the finite-size
conductance, we have strong evidence that the mode-coupling theory is
correct, and it captures the essential features of the system.

We have used parameter set E as the main data set for comparison.  The
reason for choosing this particular set of parameters is that it gives
the cleanest power-law behavior.  Other parameters will be either
qualitatively similar, or crossovers will be observed.  In fact
mode-coupling theory naturally predicts crossover, due to the presence
of $K^{(3)}$ term.  When this term is dominant, or the transverse
effect can be neglected, we should see behavior of $\kappa \propto
N^{2/5}$.  Thus we feel that the controversy of the result of Lepri {\sl
et al} \cite{2/5} and that of Narayan and Ramaswamy ($\kappa \propto
N^{1/3}$) \cite{1/3} can be reconciled within the current
mode-coupling theory.

J.-S. W. is supported by an Academic Research Grant of National
University of Singapore and by Singapore-MIT Alliance. B. L. is
supported by Academic Research Grant of National University of
Singapore.


\begin{thebibliography}{01}

\bibitem{peierls} R. E. Peierls, \textsl{Quantum Theory of Solids}, Chap.~2, (Oxford
University Press, 1955).

\bibitem{rieder-lebowitz-lieb} Z. Rieder, J. L. Lebowitz, and E. H. Lieb,
J. Math. Phys. \textbf{8}, 1073 (1967).

\bibitem{Prosen} T. Prosen and D. K. Campbell, Phys. Rev. Lett.
\textbf{84}, 2857 (2000).

\bibitem{Lebowitz} F. Bonetto, J. L. Lebowitz, L. Rey-Bellet, in \textsl{Mathematical
Physics 2000}, p.~128, A. Fokas \textsl{et al.} (eds) (Imperial
College Press, London, 2000).

\bibitem{Lepri-review} S. Lepri, R. Livi, and A. Politi, Phys. Rep.
\textbf{377}, 1 (2003).

\bibitem{2/5} S. Lepri, R. Livi, A. Politi, Europhys. Lett. \textbf{43},
271 (1998).

\bibitem{pomeau} Y. Pomeau and P. R\'esibois, Phys. Rep. \textbf{19},
63 (1975).

\bibitem{balucani} U. Balucani and M. Zoppi, \textsl{Dynamics of the Liquid State}, 
(Oxford, 1994).

\bibitem{lepri-mode-coupling} S. Lepri, Phys. Rev. E \textbf{58}, 
7165 (1998).

\bibitem{perverzev} A. Perverzev, Phys. Rev. E \textbf{68}, 056124 (2003).

\bibitem{FPUexp-Kaburaki} H. Kaburaki and M. Machida,
Phys. Lett. A\textbf{181}, 85 (1993). 

\bibitem{FPUexp-Lepri-PRL} S. Lepri, R. Livi, and A. Politi, 
Phys. Rev. Lett. \textbf{78}, 1896 (1997). 

\bibitem{FPUexp-Lepri} S. Lepri, Eur. Phys. J. B \textbf{18}, 441 (2000).

\bibitem{FPUexp-Hu} B. Hu, B. Li, and H. Zhao,
Phys. Rev. E \textbf{61}, 3828 (2000).

\bibitem{1/3} O. Narayan and S. Ramaswamy, Phys. Rev. Lett.
\textbf{89}, 200601 (2002).

\bibitem{deutsch-narayan} J. M. Deutsch and O. Narayan,
Phys. Rev. E \textbf{68}, 010201(R) (2003); 
Phys. Rev. E \textbf{68}, 041203 (2003).

\bibitem{GrassbergerYang} P. Grassberger, W. Nadler and L. Yang,
Phys. Rev. Lett. \textbf{89}, 180601 (2002).

\bibitem{Casati0} A. Dhar, Phys. Rev. Lett., {\bf
86}, 3554 (2001); G. Casati and T. Prosen,
Phys. Rev. E \textbf{67}, 015203 (R) (2003). 

\bibitem{lepri-recent} S. Lepri, R. Livi, A. Politi, Phys. Rev. E
\textbf{68}, 067102 (2003).


\bibitem{Li98} B. Hu, B. Li and H. Zhao, Phys. Rev. E \textbf{57}, 2992 (1998).

\bibitem{phi4} 
K. Aoki and D. Kusnezov, Phys. Lett. B {\bf 477}, 348 (2000).


\bibitem{mixing} B. Li, L. Wang, and B. Hu, Phys. Rev. Lett. \textbf{88}, 223901
(2002); B. Li, G. Casati, and J. Wang, Phys. Rev. E \textbf{67}, 021204
(2003); B. Li, G. Casati, J. Wang, and T. Prosen, cond-mat/0307692; D. Alonso, R. Artuso, G. Casati, and I. Guarneri
Phys. Rev. Lett. \textbf{82}, 1859 (1999);  D. Alonso, A. Ruiz, and I. de. Vega, Phys. Rev. E \textbf{66}, 066131 (2002). 

\bibitem{LiWang} B. Li and J. Wang, Phys. Rev. Lett. \textbf{91},
044301 (2003); \textsl{ibid.}, \textbf{92}, 089402 (2004).

\bibitem{diode} M. Terraneo, M. Peyrard, and G. Casati, Phys. Rev. Lett
\textbf{88}, 094302 (2002).

\bibitem{carbon-nanotube-kappa-1} S. Berber, Y.-K. Kwon, and D. Tom\'anek,
Phys. Rev. Lett. \textbf{84}, 4613 (2000).

\bibitem{carbon-nanotube-kappa-2} J. Che, T. \c{C}a\u{g}in, and W. A. Goddard III,
Nanotechnology, \textbf{11}, 65 (2000).

\bibitem{Nanotube} S. Maruyama, Physica B \textbf{323}, 193 (2002).

\bibitem{NUSpaper} Z. Yao, J.-S. Wang, B. Li, and G.-R. Liu, cond-mat/0402616; G. Zhang and B. Li, NUS-report (2004).

\bibitem{quantum-kappa} Q. Zheng, G. Su, J. Wang, and H. Guo,
Eur. Phys. J. B \textbf{25}, 233 (2002).

\bibitem{quantum-kappa-Japan} T. Yamamoto, S. Watanabe, K. Watanabe,
cond-mat/0312600. 

\bibitem{Wang-Li} J.-S. Wang and B. Li, 
Phys. Rev. Lett. \textbf{92}, 074302 (2004). 

\bibitem{Manevitch} L. I. Manevitch and A. V. Savin, Phys. Rev. E
\textbf{55}, 4713 (1997); A. V. Savin and L. I. Manevitch, {\it ibid.}
\textbf{61}, 7065 (2000); A. V. Savin,  L. I. Manevitch, P. L. Christiansen, and A. V. Zolotaryuk, Physics-Uspekhi
\textbf{42}, 245 (1999).

\bibitem{goddard} S. L. Mayo, B. D. Olafson, and W. A. Goddard III,
J. Phys. Chem. \textbf{94}, 8897 (1990).

\bibitem{tersof} J. Tersoff, Phys. Rev. B \textbf{38}, 6991 (1988).

\bibitem{Kubo} R. Kubo, M. Toda, and N. Hashitsume, \textsl{Statistical
Physics II}, pp.97-108, 2nd ed. (Springer, Berlin, 1992).

\bibitem{Zwanzig} R. Zwanzig, J. Chem. Phys. \textbf{33}, 1338 (1960).

\bibitem{Mori} H. Mori, Prog. Theor. Phys. \textbf{33}, 424 (1965).

\bibitem{Huang} K. Huang, \textsl{Statistical Mechanics}, 2nd, John
Wiley, 1987.

\bibitem{scheipers} J. Scheipers and W. Schirmacher, 
Z. Phys. B. \textbf{103}, 547 (1997). 

\bibitem{buckingham} E. Buckingham, Phys. Rev. \textbf{4}, 345 (1914).

\bibitem{frenkel-smit} For general aspects on molecular dynamics, 
see, e.g., D. Frenkel and B. Smit, \textsl{Understanding Molecular
Simulation: from algorithms to applications}, (Academic Press, 1996).

\bibitem{hardy} R. J. Hardy, Phys. Rev. \textbf{132}, 168 (1963).

\bibitem{Li-Han-Wang} B. Li, J.-H. Lan, and L. Wang, unpublished.

\end{thebibliography}
\end{document}